\documentclass[12pt,a4paper]{article}
\usepackage{axodraw}
\usepackage{cite}

\def\bea{\begin{eqnarray}}  
\def\eea{\end{eqnarray}}  
\def\bc{\begin{center}}
\def\ec{\end{center}}

\begin{document}
\pagestyle{empty}
\begin{flushright}
IFT-9/2006\\
{\tt hep-ph/0605302}\\
{\bf \today}
\end{flushright}
\vspace*{5mm}
\begin{center}

{\large {\bf (Non)decoupling of the Higgs triplet effects}}\\
\vspace*{1cm}

{\bf Piotr~H.~Chankowski}, {\bf Stefan~Pokorski} and {\bf Jakub Wagner}
\vspace{0.5cm}
 
Institute of Theoretical Physics, Warsaw University, Ho\.za 69, 00-681,
Warsaw, Poland

\vspace*{1.7cm}
{\bf Abstract}
\end{center}
\vspace*{5mm}
\noindent
{We consider the electroweak theory with an additional Higgs triplet at one 
loop using the hybrid renormalization scheme based on $\alpha_{\rm EM}$, 
$G_F$ and $M_Z$ as input observables. We show that in this scheme loop 
corrections can be naturally split into the Standard Model part and 
corrections due to ``new physics''. The latter, however do not decouple 
in the limit of infinite triplet mass parameter, if the triplet trilinear 
coupling to SM Higgs doublets grows along with the the triplet mass. In 
electroweak observables computed at one loop this effect can be attributed 
to radiative generation in this limit of a nonvanishing vacuum expectation 
value of the triplet. We also point out that whenever tree level expressions 
for the electroweak observables depend on vacuum expectation values of scalar 
fields other than the Standard Model Higgs doublet, tadpole contribution 
to the ``oblique'' parameter $T$ should in principle be included.  
In the Appendix the origin of nondecoupling is discussed on the
basis of symmetry principles in a simple scalar field theory.}
\vspace*{1.0cm}
\date{\today}


\vspace*{0.2cm}
 
\vfill\eject
\newpage

\setcounter{page}{1}
\pagestyle{plain}

\section{Introduction}

Extensions of the Standard Model (SM) constructed with the aim of solving 
the hierarchy problem are often based on gauge groups larger than 
$SU(2)\times U(1)$ and often lead to the presence in the low energy limit 
of Higgs fields in representations other than the doublet one. For example, 
string inspired models usually predict an extra $U(1)$ gauge group factor
resulting in an additional massive neutral vector boson. Some of the Little 
Higgs models predict also the existence of charged massive vector bosons 
and/or of a triplet of Higgs scalar fields. As a result in these models 
the custodial symmetry is explicitly broken and the $\rho$ parameter 
deviates from unity already at the tree level.

Precision electroweak data severely constrain such models. Some of the 
published constraints were derived by computing relevant electroweak 
observables at the tree level only \cite{CSHUKRMETE}. Others result from 
computing corrections due to new physics to the parameters $S$, $T$ and 
$U$ either at the tree level only \cite{PEPEPI} or by adding also one-loop 
corrections due to new physics \cite{HEPERI,FOROWH}. The last two approaches 
implicitly assume that the SM and new physics corrections can be separated 
from each other. 

However recently doubts have been expressed about the validity of such an 
approach \cite{JEG,CZGLJEZR,DAWSON}. It has been argued that if $\rho\neq1$ 
at the tree level, the whole structure of radiative corrections is changed 
\cite{JEG,CZGLJEZR} and the constraints on such models can only be worked 
out by computing observables in the complete model using a dedicated 
renormalization scheme: Electroweak sectors 
of such models depend, apart from the two $SU(2)\times U(1)$ gauge couplings 
$g_2$, $g_y$ and the Higgs doublet vacuum expectation value $v_H$, also
on additional free parameters (the gauge coupling(s) related to the extra 
$U(1)$ group, or the vacuum expectation value $v_\phi$ of the Higgs triplet)
and - as  has been argued in \cite{JEG,CZGLJEZR,DAWSON} - 
more than three input observables must be chosen 
in order to define the renormalization scheme. A simple $SU(2)\times U(1)$ 
model with a Higgs doublet and a Higgs triplet has been recently analysed 
in detail in ref.~\cite{DAWSON} using such a scheme (and fitted to the data
in ref.~\cite{CHDAKR}) based on $\alpha_{\rm EM}$, $G_F$, $M_Z$ and 
$\sin^2\theta$ as input observables. It has
been found that the $W$ mass depends on the top mass only logarithmically
(in contrast to the quadratic  dependence in the SM) and that contributions 
of the heavy Higgs scalars to electroweak observables do not decouple.
Similar effects were also reported in \cite{CDK} for the Littlest
Higgs model \cite{LH} also containing an $SU(2)$ triplet and earlier 
in \cite{CZGLJEZR} for the case of $U(1)$ extensions of the SM.

These results seemed surprising, especially in the case of $U(1)$ extensions 
of the SM: on the basis of the Appelquist-Carrazzone decoupling theorem one 
would expect that effects of a heavy $Z^\prime$ should be negligible and the 
SM structure of the loop corrections to electroweak observables should be only 
delicately perturbed. In order to clarify the issue we have reconsidered in 
\cite{CPW1} the $U(1)$ extension of the SM using a different renormalization 
scheme. We have pointed out that in fact the use of more input observables is 
not necessary and argued that it is precisely the use of {\it low energy} 
observables to fix the aditional parameters of the extended electroweak sector 
which is responsible for the lack of the Appelquist-Carrazzone 
decoupling of heavy sector effects in electroweak observables. The
renormalization scheme used in \cite{CPW1} is a hybrid combination of the 
scheme based on three input observables $\alpha_{\rm EM}$, $G_F$ and 
$M_W$ (for convenience) with the $\overline{\rm MS}$ scheme. Its advantage 
is that it can be applied uniformly to both, the SM and its extension. This 
enables direct comparison of the extended model predictions for various
observables (other than the input ones) with the predictions of the SM
and asessment whether they become identical in the limit of infinitely
heavy additional particles predicted by the extended model (i.e. whether
there is decoupling or not). 
We have analysed the model with the extra $U(1)$ symmetry at one loop 
using this scheme and showed that the Appelquist-Carrazzone decoupling
of a heavy $Z^\prime$ effects is manifest and that the radiative corrections
naturally split into the SM part plus corrections due to $Z^\prime$ which
are suppressed by its mass. In particular, $M_W^2$ depends on  the top
quark mass quadratically as in the SM. For the $U(1)$ extension of the SM
the proposed renormalization scheme justifies, therefore, combining the 
electroweak observables computed in the SM with higher orders corrections
included with only the tree level corrections 
due to the extended gauge structure.

In this paper we apply our renormalization scheme to the model with an
additional triplet analysed in \cite{DAWSON,CHDAKR} and earlier in 
\cite{BLHO}. Our motivation is to check whether 
the reported results are not due to the renormalization scheme adopted in 
\cite{DAWSON}. In our scheme we determine the two gauge couplings $g_2$, $g_y$ 
and the expectation value (VEV) $v_H$ of the doublet  in terms of 
$\alpha_{\rm EM}$, $G_F$ and $M_Z$ (as is customary in the SM) and treat the 
tiplet VEV $v_\phi$ as the running $\overline{\rm MS}$ parameter (at some 
fixed but arbitrary renormalization scale $Q$). The electroweak observables
computed at one loop are independent of the chosen renormalization scale 
$Q$ due to the renormalization group running of the tree level $v_\phi$
provided the tadpole contributions to the vector boson self energies
are properly included. In this scheme the loop corrections are consistently 
organized in such a way that for $v_\phi\rightarrow0$ the SM part can be 
separated from the new physics corrections. In particular, $M^2_W$ and 
$\rho$ parameters depend quadratically on the top quark mass as in the SM. 

Our renormalization scheme allows us to clarify the issue of (non)decoupling. 
As we show, at the tree level effects of the triplet in electroweak 
observables can be made arbitrarily small by decreasing the triplet VEV 
$v_\phi$ in the limit in which the triplet mass parameter $m^2_\phi$ becomes 
large. In this limit the additional neutral $K^0$ and charged $H^\pm$ scalars 
become heavy. At one loop, however, the decoupling of the triplet effects 
can be spoiled (as found in \cite{DAWSON}). This happens if the dimensionful 
coupling of two Higgs doublets with the triplet is kept of the same order as 
$m_\phi$. In our approach the violation of the decoupling in electroweak
observables is at one loop due to the 
tadpole contributions and can be attributed to a nonvanishing effective 
triplet VEV: while all additional one-particle irreducible contributions to 
electrowek observables vanish in the limit $m^2_\phi\rightarrow\infty$, the 
tadpole contributions do not: for generic values of the other parameters a 
nonzero triplet VEV is generated by radiative corrections in this limit and 
adds to the (vanishing in the $m^2_\phi\rightarrow\infty$ limit) tree level 
VEV $v_\phi$. Therefore, in this limit the electroweak obsevables are modified 
as if there was a nonzero triplet VEV in the tree level contributions. Of 
course, one can always assume that the values of the other parameters are 
such that for actual value of $m^2_\phi$ tadpole contributions vanish but 
this requires a severe fine tuning.\footnote{Beynd one loop canceling 
tadpoles by tuning parameters is also possible but there are also 
other contributions of heavy particles which do not decouple (e.g. to
the vertex $H^+W^-Z^0$ mentioned at the end of Sec. 5) and it is not
likely that simultaneous canceling of all heavy particle contributions to 
all electroweak observables is possible.} 
The tadpole contributions to the triplet VEV 
vanish in the limit $m^2_\phi\rightarrow\infty$ only if the the triplet 
dimensionfull trilinear coupling is kept of order of the electroweak 
scale (this is radiatively stable).

We also elucidate the lack of quadratic dependence of $M_W^2$ on the top 
quark mass $m_t$ reported in \cite{CZGLJEZR,DAWSON,CDK,CHDAKR}. It is simply
related to the fact that  the $W$ mass depends on $m_t$ only logarithmically 
also in the SM, if $\alpha_{\rm EM}$, $G_F$ and $\sin^2\theta^{\rm eff}_\ell$ 
are used as the input observables instead of $\alpha_{\rm EM}$, $G_F$ and 
$M_Z$. (In the SM quadratically on  $m_t$ depends then $M_Z^2$.) If the 
decoupling of the triplet effects does hold (the trilinear triplet coupling 
does not grow with $m_\phi$) the calculation of $M^2_W$ in the 
renormalization scheme based on $\alpha_{\rm EM}$, $G_F$, $M_Z$ and 
$\sin^2\theta^{\rm eff}_\ell$ can be reinterpreted as the the calculation 
of $M^2_W$ in the SM in the renormalization 
scheme based on $\alpha_{\rm EM}$, $G_F$ and $\sin^2\theta^{\rm eff}_\ell$.

The plan is as follows. In Section 2 we present the model at
the tree level. In particular we identify the limits in which the triplet
VEV vanishes. In section 3 we define our renormalization scheme and 
demonstrate its working in section 4 by computing the low energy $\rho$
defined in terms of the neutral and charged current low energy processes. 
We discuss here the (non)decoupling of the triplet effects.
In section 5 we discuss the calculation of the $W$ mass and compare it with 
the calculation in the renormalization scheme of ref \cite{DAWSON}. Finally, 
in the last section we discuss the role of tadpoles in the Peskin-Takeuchi
$S$, $T$ and $U$ parameters \cite{PETA} and comment on the problem generated 
by the presence of $SU(2)$ Higgs triplets in Grand Unified Theories.
{\bf Since the nondecoupling of Higgs triplet effects may seem 
counter-intuitive, and has important consequences for model building,
in \ref{app:nondecoupling} we investigate it from another point of view 
in the model with Higgs fields only. 
\ref{app:formulae} contains necessary formulae.} 

\section{The model}

As in \cite{DAWSON,FOROWH} we consider for simplicity the SM model extended 
with a $Y=0$ Higgs (weak hypercharge) triplet $\Phi={1\over2}\tau^a\phi^a$.
We would like to investigate whether the triplet can decouple, that is whether 
there is a limit in which the model predictions for all ``low energy'' 
observables (i.e. observables which can be defined also in the SM) are the 
same as in the SM.

Assuming that both, the doublet and the triplet acquire vacuum expectation
values (VEVs) $v_H$ and $v_\phi$, respectively, the tree level expressions for 
gauge boson masses read
\begin{eqnarray}
M^2_Z&=&{1\over4}(g^2_2+g^2_y)v^2_H~,\nonumber\\
M^2_W&=&{1\over4}g^2_2 v^2_H+g^2_2 v^2_\phi~.
\label{eqn:gaugebosmasses}
\end{eqnarray}
The gauge boson couplings to fermions are as in the SM (similarly as the 
Yukawa terms). The weak mixing angle is also defined by 
$s^2=g^2_y/(g^2_2+g^2_y)$. The three customary SM input 
observables are given by
\begin{eqnarray}
M^2_Z&=&{1\over4}{e^2\over s^2 c^2} v_H^2~,\nonumber\\
\alpha_{\rm EM} &=& e^2/4\pi~,\label{eqn:basicreltree}\\
\sqrt2 G_F&=&{e^2\over 4s^2M^2_W}
={1\over v^2_H+4v^2_\phi}~.\nonumber
\end{eqnarray}
so that
\begin{eqnarray}
e^2&=&4\pi\alpha_{\rm EM}\equiv e^2_{(0)}~,\nonumber\\
v^2_H&=&{1\over\sqrt2G_F}-4v^2_\phi~,\label{eqn:inversebasicreltree}\\
s^2&=&{1\over2}\left(1-\sqrt{1-{4\pi\alpha_{\rm EM}\over\sqrt2G_FM^2_Z}
\left(1-4\sqrt2G_Fv^2_\phi\right)}\phantom{i}\right)\equiv s^2_{(0)}~.
\nonumber
\end{eqnarray}
In terms of $M^2_Z$, $\alpha_{\rm EM}$ and $G_F$ the $W$ boson mass is 
then given by 
\begin{eqnarray}
M_W^2={\pi\alpha_{\rm EM}\over\sqrt2G_Fs^2_{(0)}}\equiv (M_W^2)_{(0)}~.
\label{eqn:Mwzerothorder}
\end{eqnarray}
and all measurable $\rho$ parameters that can be defined 
\cite{CPW1} are equal:
\begin{eqnarray}
\rho_{\rm low}=\rho=\rho_{Zf}={1\over1-4\sqrt2G_Fv^2_\phi}~.
\label{eqn:rhostree}
\end{eqnarray}
It is therefore clear that the first condition for decoupling, 
i.e. for decoupling at the tree level, of the 
triplet is vanishing (in the limit) of its tree level VEV $v_\phi$.

The tree level VEVs $v_\phi$ and $v_H$ 
\begin{eqnarray}
\langle H\rangle={1\over\sqrt2}\left(\matrix{0\cr v_H}\right)~,
\phantom{aaaaaaaa}
\langle\phi^0\rangle=v_\phi\label{eqn:VEVs}
\end{eqnarray}
are determined by the Higgs potential whose most general form is
\begin{eqnarray}
V=m_H^2H^\dagger H+{\lambda_H\over4}\left(H^\dagger H\right)^2
+m^2_\phi~{\rm tr}\left(\Phi^2\right)
+{\lambda_\phi\over4}~\left({\rm tr}\Phi^2\right)^2\nonumber\\
+\kappa ~H^\dagger H~{\rm tr}\left(\Phi^2\right)+\mu ~H^\dagger\Phi H
\phantom{aaaaaaaaaaaaaa}\label{eqn:potential}
\end{eqnarray}
where tr$\left(\Phi^2\right)={1\over2}\left(\phi^0\phi^0+2\phi^+\phi^-\right)$.
The minimization conditions determining $v_\phi$ and $v_H$ read
\begin{eqnarray}
v_H\left(m_H^2+{\lambda_H\over4}v_H^2
+{\kappa\over2}v_\phi^2-{1\over2}\mu v_\phi\right)=0
\nonumber\\
m^2_\phi v_\phi+{\lambda_\phi\over4}v_\phi^3+{\kappa\over2}v_H^2v_\phi
-{1\over4}\mu v_H^2=0\label{eqn:minimizconds}
\end{eqnarray}
There are two limits in which $v_\phi\rightarrow0$: 
\begin{itemize}
\item For $\mu=0$ and for $m^2_\phi$ and $\kappa$ positive ($\lambda_\phi$ 
must be positive for stability anyway) the solution to the second condition 
is clearly $v_\phi=0$ (and evidently $v_\phi\neq0$ for $\mu\neq0$ if $v_H$
is nonzero). It is therefore easy to see that $v_\phi\propto\mu$ for 
$\mu\rightarrow0$, independently of the magnitude of $m^2_\phi$. 
\item For positive $m^2_\phi\rightarrow\infty$ and $\lambda_\phi$, $\kappa$ 
fixed the first term in the second equation of (\ref{eqn:minimizconds}) 
cannot be canceled by the other terms unless 
$v_\phi\rightarrow0$. In the limit $v_\phi\sim{1\over4}\mu (v_H^2/m^2_\phi)$. 
\end{itemize}
Thus, in both these limits the effects of the triplet in low energy
electroweak observables vanish at the tree level. 

To investigate whether the triplet decouples also with quantum corrections 
taken into account we first determine the masses of the physical scalars.
They are given by 
\begin{eqnarray}
V_{\rm quadr}&=&\left(\mu v_\phi+{v_H^2\over4}{\mu\over v_\phi}\right)
H^+H^-\nonumber\\
&+&{1\over2}\left(h^0,\phi^0\right)
\left(\matrix{{1\over2}\lambda_Hv^2_H&(\kappa v_\phi-{1\over2}\mu)v_H\cr
(\kappa v_\phi-{1\over2}\mu)v_H&
{1\over2}\lambda_\phi v^2_\phi+{v^2_H\over4}{\mu\over v_\phi} }\right)
\left(\matrix{h^0\cr\phi^0}\right)\label{eqn:massmatrices}
\end{eqnarray}
The physical charged scalar $H^+$ and $G^+_W$ which becomes the 
longitudinal component of the massive $W^\pm$ are given by the following 
combinations of the charged doublet ($G^+$) and triplet ($\phi^+$) components
\begin{eqnarray}
G^+_W=c_\delta G^+-s_\delta\phi^+\phantom{aaaaaaaa}
G^+=c_\delta G^+_W+s_\delta H^+~,
\nonumber\\
H^+=s_\delta G^++c_\delta\phi^+\phantom{aaaaaaaa}
\phi^+=-s_\delta G^+_W+c_\delta H^+~,\label{eqn:chargedHiggsrot}
\end{eqnarray}
where 
\begin{eqnarray}
s_\delta=2{v_\phi\over v_H}{1\over\sqrt{1+4v^2_\phi/v^2_H}}~,
\phantom{aaaa}
c_\delta={1\over\sqrt{1+4v^2_\phi/v^2_H}}~,\label{eqn:scdelta}
\end{eqnarray}
The neutral mass eigenstates $H^0$ and $K^0$ are given by
\begin{eqnarray}
\left(\matrix{h^0\cr\phi^0}\right)=\left(\matrix{c_\gamma&-s_\gamma\cr
s_\gamma&c_\gamma}\right)\left(\matrix{H^0\cr K^0}\right)~.
\label{eqn:neutralHiggsrot}
\end{eqnarray}
Useful relations between $c_\gamma$, $s_\gamma$ and $H^0$ and $K^0$ masses 
are given in \ref{app:formulae}. Here we note only that for 
$v_\phi\rightarrow0$ the mixing between $G^+$ and 
$\phi^+$ as well as the mixing between $h^0$ and $\phi^0$ vanishes:
For  $v_\phi\rightarrow0$ one has
\begin{eqnarray}
s_\delta\approx s_\gamma\approx  {\mu\over2}{v_H\over m^2_\phi}\rightarrow0~.
\label{eqn:sinussesinthelimit}
\end{eqnarray}

From the formulae given above it is clear that for $\mu\rightarrow0$ the 
masses of $H^+$ and of $K^0$ approach $M_{H^+}=M_{K^0}\sim m_\phi$. Hence 
although $v_\phi\rightarrow0$ in this limit, loop contributions of $H^+$ and 
$K^0$ to electroweak observables will not be suppressed if $m_\phi$ is not 
large, and decoupling cannot hold. In contrast, for 
$m^2_\phi\rightarrow\infty$, $M_{H^+}=M_{K^0}\rightarrow m_\phi$ 
and decoupling can be expected. However, as we will show, it holds
only in the limit $m^2_\phi\rightarrow\infty$, $\mu$ fixed. In the
limit $\mu\propto m_\phi\rightarrow\infty$ it is broken by quantum corrections
and unless the other parameters are tuned appropriately the effects of $H^+$ 
and $K^0$ cannot be neglected.

\section{Renormalization scheme}
\label {sec:scheme}

Beyond the tree level the model can be most easily renormalized using the 
minimal subtraction (at some fixed renormalization scale), so that its 
predictions for observables are given as (finite) functions of the 
renormalized running Lagrangian parameters. It is customary to invert 
the appropriate number of these relations and to express all
running parameters in terms of a chosen set of input observables
(the number of the input observables must be then equal to the number of
renormalized parameters), so that other observables are predicted in terms of
the chosen ones. This step is however not necessary. In principle it is
perfectly possible to fit the renormalized parameters to the data directly.
It is also possible, as proposed in \cite{CPW1}, to invert only a smaller 
set of relations and to express predictions of the model in terms of
some smaller number (smaller than the number of renormalized parameters) 
of input observables and a complementary number of renormalized parameters.

Following this logic, if the SM is naturally embedded in an extended model, 
the same {\it three} ``low energy'' input observables can be chosen for the SM 
and its extension and used to eliminate the {\it same} three combinations of 
the renormalized parameters (e.g. $g_y$, $g_2$ and $v_H$) in both models 
keeping additional parameters of the extended model as renormalized free 
parameters. This enables direct comparison of the extended model 
predictions for other observables  with the predictions of the SM because  
the structure of radiative corrections in the extended model is such that the 
SM contributions can be clearly separated. The decoupling of heavy extra 
particles cannot be then superficially spoiled and becomes easy to assess: it
holds if in the limit of infinitely heavy additional particles (not present 
in the SM) predictions of the extended model for observables (other than the 
input ones) become identical with the predictions of the SM. In fact this is 
almost the unique way of checking the decoupling. A possible modification 
would be to determine additional parameters of the extended model using 
observables which do not exist in the SM (``high energy observables'') 
such as physical masses and couplings of the additional particles. This 
step is however an unnecessary technical complication. 

Renormalization schemes which use different numbers of ``low energy'' input 
observables for the SM and for its extension do not allow for checking 
decoupling directly. Using more measured ``low energy'' input 
observables to fix other parameters of the extended model usually spoils 
natural correlations 
(existing in the extended theory) of the values of these parameters with 
masses of the heavy particles which are necessary for the decoupling to hold. 
Such schemes can, therefore, spoil decoupling superficially. This was 
demonstrated in \cite{CPW1} in the case of a $U(1)$ extension in which
fixing the $Z^\prime$ gauge coupling in terms of an additional low energy
observable (e.g. $\sin^2\theta_\ell^{\rm eff}$) disables in fact taking 
$M_{Z^\prime}$ to infinity, if the running coupling constant is not to 
become nonperturbatively large.

In order to avoid such complications  we shall analyse the triplet extension 
of the SM using for the input observables $\alpha_{\rm EM}$, $G_F$, $M_Z$ as 
in the SM. The model has 4 free parameters: the gauge couplings $g_y$, $g_2$ 
and the VEVs $v_H$ and $v_\phi$. Beyond the tree level they are interpreted as 
the renormalized running parameters. For given $\hat v_\phi$ we express 
$\hat g_y$, $\hat g_2$ and $\hat v_H$ (from now on we denote the Lagrangian 
running parameters by a hat) in terms of $\alpha_{\rm EM}$ $G_F$, $M_Z$. Other 
physical quantities are then given as functions of $\alpha_{\rm EM}$, $G_F$, 
$M_Z$ and $\hat v_\phi$ which in reality is a function of the running 
parameters of the Higgs potential (\ref{eqn:potential}). The relations which 
for this purpose must be inverted are
\begin{eqnarray}
M^2_Z&=&{1\over4}{\hat e^2\over\hat s^2\hat c^2}\hat v_H^2
+\hat\Pi_{ZZ}(M^2_Z)\nonumber\\
\alpha_{\rm EM} &=& 
\hat\alpha\left(1+\tilde\Pi_\gamma(0)-{\hat\alpha\over\pi}\ln
{\hat M^2_W\over Q^2}\right)\label{eqn:basicrelsatoneloop}\\
\sqrt2 G_F&=&{1\over\hat v^2_H+4\hat v^2_\phi}\left(1+\Delta_G\right)
\nonumber
\end{eqnarray}
where 
\begin{eqnarray}
\tilde\Pi_\gamma(0)=\hat e^2\left[{2\over3}-3\ln{\hat M^2_W\over Q^2}
+{4\over3}\sum_fN_c^{(f)}Q_f^2\ln{\hat m^2_f\over Q^2}
+{1\over3}\ln{\hat M^2_{H^+}\over Q^2}\right]
\end{eqnarray}
\begin{eqnarray}
\Delta_G = 
-{\hat\Pi_{WW}(0)\over\hat M^2_W}+B_{W\gamma}+B_{WZ^0}
+2\hat\Lambda+\hat\Sigma_{eL}+\hat\Sigma_{\nu L}~.\label{eqn:DeltaG}
\end{eqnarray}
Explicit expressions for $\Delta_G$ are given in \ref{app:formulae}.
The last term in the expression for $\alpha_{\rm EM}$ can be computed along the
lines given in \cite{POK,CPW1}.

Solving for $\hat\alpha$, $\hat s$ and $\hat v_H$ with one loop accuracy 
we find
\begin{eqnarray}
\hat\alpha&=&\alpha_{\rm EM}
\left(1-\tilde\Pi_\gamma(0)+{\alpha_{\rm EM}\over\pi}
\ln{(M^2_W)_{(0)}\over Q^2}\right)\nonumber\\
\hat v^2_H&=&{1\over\sqrt2G_F}(1+\Delta_G)-4\hat v^2_\phi\label{eqn:GFcorr1}\\
\hat s^2\hat c^2&=&{\pi\alpha_{\rm EM}\over\sqrt2G_FM^2_Z}
\left(1-\tilde\Pi_\gamma(0)+{\alpha_{\rm EM}\over\pi}
\ln{(M^2_W)_{(0)}\over Q^2}+{\hat\Pi_{ZZ}(M_Z^2)\over M_Z^2}\right)
\left[1-4\sqrt2G_F\hat v^2_\phi+\Delta_G\right]\nonumber
\end{eqnarray}
where $(M^2_W)_{(0)}$ is defined in (\ref{eqn:Mwzerothorder}).
The last relation gives for $\hat s^2$
\begin{eqnarray}
\hat s^2={1\over2}\left(1-\sqrt{
1-{4\pi\alpha_{\rm EM}\over\sqrt2G_FM^2_Z}
\left(1-4\sqrt2G_F\hat v^2_\phi+\Delta\right)}\phantom{i}\right)
\label{eqn:shatatoneloop}
\end{eqnarray}
where
\begin{eqnarray}
\Delta=\Delta_G-\left(1-4\sqrt2G_F\hat v^2_\phi\right)
\left(\tilde\Pi_\gamma(0)-{\alpha_{\rm EM}\over\pi}
\ln{(M^2_W)_{(0)}\over Q^2}-{\hat\Pi_{ZZ}(M_Z^2)\over M_Z^2}\right)
\label{eqn:Deltadefinition}
\end{eqnarray}
so that to one loop accuracy
\begin{eqnarray}
\hat s^2=s^2_{(0)}+{s^2_{(0)}c^2_{(0)}\over c^2_{(0)}-s^2_{(0)}}
\left[{\Delta_G\over1-4\sqrt2G_F\hat v^2_\phi}
-\tilde\Pi_\gamma(0)+{\alpha_{\rm EM}\over\pi}
\ln{M^2_W\over Q^2}+{\hat\Pi_{ZZ}(M_Z^2)\over M_Z^2}\right]\phantom{a}
\label{eqn:s2tooneloopac}
\end{eqnarray}
where $s^2_{(0)}$ is defined in (\ref{eqn:inversebasicreltree}).
The formulae (\ref{eqn:GFcorr1}) and (\ref{eqn:s2tooneloopac}) form
the basis of our renormalization scheme: 
they allow to express $\hat\alpha$, $\hat v^2_H$, $\hat s^2$ and 
$\hat c^2$ in formulae for electroweak observables in terms of input
observables to one loop accuracy.

\section{$\rho_{\rm low}$ in the electroweak model with $Y=0$ triplet}

In order to check decoupling of the triplet we compute at one loop
the parameter $\rho_{\rm low}$ defined in terms of the ratio of the 
neutral and charged current terms in the low energy effective Lagrangian. 
It is given by 
\begin{eqnarray}
\rho_{\rm low}= {a-b\over\sqrt2G_F}~,
\end{eqnarray}
where $a$ and $b$ are the coefficients in the neutral current low energy 
effective Lagrangian written in the form
\begin{eqnarray}
{\cal L}^{\rm eff}=
[\bar\psi_e\gamma^\lambda(a\mathbf{P}_L+b\mathbf{P}_R)\psi_e]
[\bar\psi_{\nu_\mu}\gamma_\lambda\mathbf{P}_L\psi_{\nu_\mu}]
\label{eqn:rhodefinition2}
\end{eqnarray}
In terms of the running Lagrangian parameters we get
\begin{eqnarray}
\sqrt2G_F\rho_{\rm low}={\hat e^2\over4\hat s^2\hat c^2\hat M^2_Z}\left\{
1-{\hat\Pi_{ZZ}(0)\over\hat M^2_Z}
-{\hat e^2\over4\pi^2}{\hat c^2\over\hat s^2}\ln{\hat M^2_W\over Q^2}
\right.\phantom{aaaaaaaaaa}\nonumber\\
\left.+
{\hat e^2\over64\pi^2\hat s^2\hat c^2}\left(3-12\hat s^2+24\hat s^4
+16\hat c^4{\hat M^2_Z\over\hat M^2_W}\right)\right\}
\end{eqnarray}
where the last two terms are the contributions of the vertex and box diagram 
corrections, respectively. These are computed as in the SM except that one 
cannot use the relation $\hat M^2_W=\hat c^2\hat M^2_Z$. To one loop accuracy, 
using the relations (\ref{eqn:GFcorr1}) and (\ref{eqn:Mwzerothorder}) we have 
therefore
\begin{eqnarray}
\sqrt2G_F\rho_{\rm low}={\sqrt2G_F\over1-4\sqrt2G_F\hat v_\phi^2}\left\{
1-{\Delta_G\over1-4\sqrt2G_F\hat v_\phi^2}-{\hat\Pi_{ZZ}(0)\over M^2_Z}\right.
-{e^2_{(0)}\over4\pi^2}
{c^2_{(0)}\over s^2_{(0)}}\ln{(M^2_W)_{(0)}\over Q^2}
\phantom{aaa}\nonumber\\
\left.
+{e^2_{(0)}\over64\pi^2s^2_{(0)}c^2_{(0)}}\left(3-12s^2_{(0)}+24s^4_{(0)}
+16c^4_{(0)}{M^2_Z\over(M^2_W)_{(0)}}\right)\right\}
\label{eqn:rholowgeneral}
\end{eqnarray}
One has to remember that the $Z^0$ and $W$ self energies  (in $\Delta_G$)
include tadpole contributions. Note also that in agreement with
(\ref{eqn:rhostree}) to one loop
\begin{eqnarray}
{1\over1-4\sqrt2G_F\hat v_\phi^2}{\hat\Pi_{WW}(0)\over(M^2_W)_{(0)}}
-{\hat\Pi_{ZZ}(0)\over M^2_Z}
={\hat\Pi_{WW}(0)\over c^2_{(0)}M^2_Z}-{\hat\Pi_{ZZ}(0)\over M^2_Z}+\dots
\end{eqnarray}

We can now examine different contributions to $\rho$.
Since the coupling of fermions to gauge bosons are as in the SM
for the one particle irreducible contribution of top and bottom
to $\rho_{\rm low}$ we get
\begin{eqnarray}
\rho_{\rm low}={1\over1-4\sqrt2G_F\hat v^2_\phi}\left(1+
{\sqrt2G_F\over1-4\sqrt2G_F\hat v^2_\phi}~{N_c\over16\pi^2}~g(m_t,m_b)
+\dots\right)\label{eqn:tbtorholow}
\end{eqnarray}
where $g(m_t,m_b)\approx m^2_t$ is defined in \ref{app:formulae}.
This is finite as in the SM. Moreover, it is clear that in the limit 
$m^2_\phi\rightarrow\infty$, when $\hat v_\phi\rightarrow0$, the expression
(\ref{eqn:tbtorholow}) 
reduces to the well known SM expression. Fermions contribute to $\rho$ also 
through the tadpoles, but this second contribution vanishes as 
$\hat v_\phi\rightarrow0$. 

The full expressions for $\hat\Pi_{WW}(q^2)$ and $\hat\Pi_{ZZ}(q^2)$
are given in \ref{app:formulae}. It is easy to see that the limit
$m^2_\phi\rightarrow\infty$ ($s_\gamma$, $s_\delta\rightarrow0$) 
gauge boson contributions and most of the 1PI contributions of scalars 
approach the SM limit and one is left with the following dangerous, 
because\footnote{Note that for $m^2_\phi\rightarrow\infty$
$\tilde A(0,M_{K^0},M_{H^+})\rightarrow0$.} 
$\tilde A(0,M_{Z,W,H^0},M_{H^+,K^0})\rightarrow-{1\over8}m^2_\phi$, terms
\begin{eqnarray}
{\hat\Pi_{WW}(0)\over\hat c^2}-\hat\Pi_{ZZ}(0)&\supset&
{\hat e^2\over\hat s^2\hat c^2}\left[
-4\left(-s_\gamma c_\delta+{1\over2}c_\gamma s_\delta\right)^2
\tilde A(0,M_{H^0},M_{H^+})\right.\nonumber\\
&&\phantom{aaaa}
-4\left(c_\gamma s_\delta-{1\over2}s_\gamma c_\delta\right)^2
\tilde A(0,M_W,M_{K^0})\label{eqn:dangerous}\\
&&\phantom{aaa}\left.\matrix{\phantom{a}\cr\phantom{a}}
-s_\delta^2\tilde A(0,M_Z,M_{H^+})\right]\nonumber\\
&&-{\hat e^2\over\hat s^2\hat c^2}\left[
-2s_\delta^2c_\delta^2\tilde A(0,M_W,M_{H^+})-s_\gamma^2
\tilde A(0,M_Z,M_{K^0})\right]\nonumber
\end{eqnarray}
The other contributions, in particular, those with $b_0$ functions
are suppressed for $m^2_\phi\rightarrow\infty$. If the dimensionfull 
coupling $\mu$ in (\ref{eqn:potential}) grows along with $m_\phi$,
so that $\mu\propto m_\phi$ and $s_\gamma\approx s_\delta\sim1/m_\phi$,
each individual term in (\ref{eqn:dangerous})
approaches a constant (they vanish if $\mu$ is kept fixed). However using 
(\ref{eqn:sinussesinthelimit}) it is easy to check that together they cancel 
out, so that the 1PI cotributions of extra scalars to $\rho_{\rm low}$ does 
decouple. 

The complete 1PI bosonic contribution to $\rho_{\rm low}$ is not independent
of the renormalization scale $Q$. Another explicit dependence on  $Q$ is 
introduced by tadpole contributions to $\hat\Pi_{WW}$ and $\hat\Pi_{ZZ}$. 
Due to our renormalization scheme (in which $\hat v_\phi$ is not traded for 
an observable), the contribution of tadpoles to $\rho_{\rm low}$ (and
other observables) does not cancel out but is found to be
\begin{eqnarray}
\rho_{\rm low}
={1\over1-4\sqrt2G_F\hat v_\phi^2}\left\{1
-8{\sqrt2G_F\hat v_\phi\over1-4\sqrt2G_F\hat v_\phi^2}
\left(s_\gamma{{\cal T}_H\over\hat  M^2_{H^0}}
+c_\gamma{{\cal T}_K\over\hat  M^2_{K^0}}\right)
+\dots\right\}\label{eqn:tadpolestorho}
\end{eqnarray}
where $\langle H^0\rangle_{\rm loop}=-i{\cal T}_H$, 
$\langle K^0\rangle_{\rm loop}=-i{\cal T}_K$. 
For example fermions give
\begin{eqnarray}
{\cal T}_H&=&-2\sqrt2c_\gamma\sum_fN_c^{(f)}\hat Y_f\hat m_fa(\hat m_f)
\nonumber\\
{\cal T}_K&=&+2\sqrt2s_\gamma\sum_fN_c^{(f)}\hat Y_f\hat m_fa(\hat m_f)
\nonumber
\end{eqnarray}
where $\hat Y_f$ are the Yukawa couplings. Explicit dependence of tadpoles on 
$Q$ is necessary to render $\rho_{\rm low}$ 
renormalization scale independent \cite{CHWA,CPW1}. This is because the tree 
level expression for $\rho_{\rm low}$ depends on the running parameter 
$\hat v_\phi$ which also changes with the renormalization scale. Therefore to
one loop  accuracy
\begin{eqnarray}
\rho_{\rm low}&=&{1\over1-4\sqrt2G_F\hat v_\phi^2(Q)}
\left\{1+\dots\right\}\nonumber\\
&=&{1\over1-4\sqrt2G_F\hat v_\phi^2(Q^\prime)}
\left\{1+2{\sqrt2G_F\over1-4\sqrt2G_F\hat v_\phi^2}\dot v_\phi^2
\ln{Q^2\over Q^{\prime2}}+\dots\right\}\label{eqn:changeofscale}
\end{eqnarray}
where $\dot v_\phi^2\equiv Q d\hat v_\phi^2/dQ$.
Using the relations (\ref{eqn:inverseinverserelations}) it is for example easy 
to see that the terms $\sim\hat Y^4_f$ in the renormalization group equation 
(\ref{eqn:RGEforvphi}), when inserted in (\ref{eqn:changeofscale}), for 
$\dot v_\phi^2$ properly change $Q$ into $Q^\prime$ in the fermionic tadpoles. 
Similarly, $Q$ dependence of the gauge boson (and ghost) tadpoles 
\begin{eqnarray}
s_\gamma{{\cal T}_H^{(W,Z)}\over\hat  M^2_{H^0}}+
c_\gamma{{\cal T}_K^{(W,Z)}\over\hat  M^2_{K^0}}
&=&{1\over16\pi^2{\rm DET}}\left[
3{\hat e^2\over\hat  s^2}\hat \lambda_H\hat v^2_H\hat v_\phi
\hat M^2_W\left(\ln{\hat M^2_W\over Q^2}-{1\over3}\right)\right.\\
&&\phantom{aaaaaa}
-{3\over2}{\hat e^2\over\hat  s^2}\hat v^2_H
\left(\hat \kappa\hat  v_\phi-{1\over2}\hat \mu\right)
\hat M^2_W\left(\ln{\hat M^2_W\over Q^2}-{1\over3}\right)\nonumber\\
&&\phantom{aaaaaa}-\left. {3\over4}{\hat e^2\over\hat  s^2c^2}\hat v^2_H
\left(\hat\kappa\hat v_\phi-{1\over2}\hat\mu\right)
\hat M^2_Z\left(\ln{\hat M^2_Z\over Q^2}-{1\over3}\right)\right]
\nonumber
\end{eqnarray}
where DET is given by (\ref{eqn:DET}) combines with the terms
$\propto \hat g_2^4,\hat g_2^2\hat g_y^2,\hat g_y^4$ in 
(\ref{eqn:RGEforvphi}). 
Contribution of $G^0$ (the $Z^0$ Goldstone) to the combination of tadpoles in
(\ref{eqn:tadpolestorho}) vanishes whereas the 
renormalization scale dependence of the $G^\pm_W$ ($W^\pm$ Goldstones) 
contribution  
\begin{eqnarray}
s_\gamma{{\cal T}_H^{(G^\pm)}\over\hat M^2_{H^0}}+
c_\gamma{{\cal T}_K^{(G^\pm)}\over\hat M^2_{K^0}}
={1\over16\pi^2}{4\hat v_\phi\over\hat v^2_H+4\hat v^2_\phi}a(\hat M_W)
\end{eqnarray}
adds to the remaining $Q$ dependence of the
bosonic 1PI contributions to $\rho_{\rm low}$ and together they can be seen
to properly match the term $\propto\hat g^2_2$ in (\ref{eqn:RGEforvphi}). 

The $H^\pm$ contribution to the tadpole combination in 
(\ref{eqn:tadpolestorho}) is
\begin{eqnarray}
s_\gamma{{\cal T}_H^{(H^\pm)}\over\hat  M^2_{H^0}}+
c_\gamma{{\cal T}_K^{(H^\pm)}\over\hat  M^2_{K^0}}
={1\over16\pi^2}{1\over{\rm DET}(1+4\hat v^2_\phi/\hat v^2_H)}
\phantom{aaaaaaaaaaaaaaaaaaaaaaaaa}\nonumber\\
\times\left[2\hat \lambda_H\hat \mu\hat  v_\phi^2
+{1\over4}\hat \lambda_H\hat \lambda_\phi\hat  v_H^2\hat v_\phi
-2\hat \kappa\hat \mu\hat  v_\phi^2 -\hat \kappa^2\hat v^2_H\hat v_\phi 
+{1\over2}\hat \kappa\hat \mu v^2_H+\hat \mu^2\hat v_\phi
\right]a(\hat M_{H^+})\phantom{aaa}
\end{eqnarray}
whereas the neutral scalars contribution to the tadpole combination in 
(\ref{eqn:tadpolestorho}) reads
\begin{eqnarray}
s_\gamma{{\cal T}_H^{(H^0,K^0)}\over\hat  M^2_{H^0}}+
c_\gamma{{\cal T}_K^{(H^0,K^0)}\over\hat  M^2_{K^0}}
={1\over{\rm DET}}\left[
-{3\over4}\hat \lambda_H\hat v^2_H\left(\hat \kappa\hat  v_\phi-{1\over2}
\hat \mu\right)\left(c^2_\gamma a(\hat M_{H^0})
+s^2_\gamma a(\hat M_{K^0})\right)\right.
\nonumber\\
+{3\over8}\hat \lambda_H\hat \lambda_\phi\hat  v^2_H\hat v_\phi 
\left(s^2_\gamma a(\hat M_{H^0})+c^2_\gamma a(\hat M_{K^0})\right)
\phantom{aaaaa}\nonumber\\
+{1\over4}\hat \lambda_H\hat \kappa\hat  v_H^3
\left(a(\hat M_{H^0})- a(\hat M_{K^0})\right)
2s_\gamma c_\gamma
\phantom{aaaaaa}\nonumber\\
-{1\over2}\hat \kappa\hat  v^2_H\left(\hat \kappa\hat  v_\phi-{1\over2}
\hat \mu\right)\left(s^2_\gamma a(\hat M_{H^0})
+c^2_\gamma a(\hat M_{K^0})\right)
\phantom{a}\nonumber\\
+{1\over4}\hat \lambda_H\hat  v^2_H\left(\hat \kappa\hat  v_\phi-{1\over2}
\hat \mu\right)
\left(c^2_\gamma a(\hat M_{H^0})+s^2_\gamma a(\hat M_{K^0})\right)
\nonumber\\
\left.
-{1\over2}\hat v_H\left(\hat \kappa\hat  v_\phi-{1\over2}\hat \mu\right)^2
\left(a(\hat M_{H^0})- a(\hat M_{K^0})\right)2s_\gamma c_\gamma\right]
\nonumber
\end{eqnarray}

In the limit $m_\phi\rightarrow\infty$, $\mu\propto m_\phi$ the $K^0$ and 
$H^+$ tadpole contributions  do not vanish in general. Their contribution to 
$\rho_{\rm low}$ is in this limit given by (\ref{eqn:tadpolestorho}) with 
\begin{eqnarray}
\hat v_\phi\left(s_\gamma{{\cal T}_H^{(H^\pm,K^0)}\over\hat  M^2_{H^0}}+
c_\gamma{{\cal T}_K^{(H^\pm,K^0)}\over\hat  M^2_{K^0}}\right)
={1\over16\pi^2}{3\hat \mu\hat  v_\phi({1\over2}\hat \kappa\hat  v^2_H
+\hat \mu\hat  v_\phi)\over2\hat M^2_{H^0}}
\left(-1+\ln{\hat m^2_\phi\over Q^2}\right)
\end{eqnarray}
Hence, unless one assumes that for the particular renormalization scale 
$Q$ chosen for calculations $\ln(\hat m^2_\phi(Q)/Q^2)\approx1$ or that
$\hat\kappa(Q)\hat v^2_H(Q)+2\hat\mu(Q)\hat v_\phi(Q)\approx 0$ (these 
relations would be, of course, modified in higher orders), there is no 
decoupling in this limit. Our approach allows to understand this peculiarity 
(observed in \cite{DAWSON}) as due to the behaviour of the triplet VEV. In 
the above limit the tree level VEV $\hat v_\phi$ vanishes but the one loop 
corrections to the true VEV of the triplet do not. This is because once the 
$SU(2)$ symmetry is broken (by the VEV of the doublet), the triplet is no 
longer protected from acquiring a nonzero VEV radiatively. Thus, although 
the tree level triplet VEV, as well as all 1PI one loop contributions to 
$\rho_{\rm low}$, vanish for $\hat m_\phi\rightarrow\infty$, 
$\hat\mu\propto\hat m_\phi$, there is a nonzero correction to the SM result 
which can be accounted for by simply replacing $v_\phi$ by one loop correction 
to it in the tree level term in $\rho_{\rm low}$. Of course, if the 
dimensionful coupling $\hat \mu$ is kept fixed, the $K^0$ and $H^+$ 
contribution to tadpoles vanish in the limit $\hat m_\phi\rightarrow\infty$ 
and one recovers the SM result. {\bf In order to better understand the origin 
of nondecoupling we analyze it in \ref{app:nondecoupling} in the simplified 
model with Higgs fields only and point out its connections with the custodial
symmetry breaking.}

\section{The $W$ boson mass}

In this section we compare the one loop expression for the $W$ boson mass
in our scheme and in the scheme of ref. \cite{DAWSON}. This will allow us
to elucidate the question of its dependence on the top quark mass.

In our scheme the $W$ boson mass is at one loop given by 
\begin{eqnarray}
M^2_W = {\hat e^2\over4\hat s^2}(\hat v_H^2+4\hat v^2_\phi)+\hat\Pi_{WW}(M^2_W)
\end{eqnarray}
Using the one loop expressions for $\hat e^2$, $\hat s^2$ and $\hat v^2_H$
(\ref{eqn:GFcorr1}), (\ref{eqn:shatatoneloop}) and  
(\ref{eqn:Deltadefinition}) this takes the form
\begin{eqnarray}
M^2_W = {\pi\alpha_{\rm EM}\over\sqrt2G_Fs^2_{(0)}}\left\{
1-\tilde\Pi_\gamma(0)+{\alpha_{\rm EM}\over\pi}\ln{(M^2_W)_{(0)}\over Q^2}
+\Delta_G +{\hat\Pi_{WW}((M^2_W)_{(0)})\over(M^2_W)_{(0)}}\right.
\phantom{aaaaa}\label{eqn:Wmass}\\
-{c^2_{(0)}\over c^2_{(0)}-s^2_{(0)}}
\left[{\hat\Pi_{ZZ}(M^2_Z)\over M^2_Z}-{\hat\Pi_{WW}(0)\over c^2_{(0)}M^2_Z}
-\tilde\Pi_\gamma(0)+{\alpha_{\rm EM}\over\pi}\ln{(M^2_W)_{(0)}\over Q^2}
\right.\phantom{a}\nonumber\\
\left.\left.
+{1\over1-4\sqrt2G_F\hat v^2_\phi}
\left(B_{W\gamma}+B_{WZ^0}+2\hat\Lambda
+\hat\Sigma_{eL}+\hat\Sigma_{\nu L}\right)\right]\right\}\nonumber
\end{eqnarray}
The non-tadpole fermion contribution to this formula has formally exactly 
the same form as in the SM. Hence, it is finite and renormalization  scale 
independent. Expressed in terms of the input observables $\alpha_{\rm EM}$, 
$G_F$ and $M_Z$ it differs, however, from the corresponding contribution in 
the SM in that $s^2_{(0)}$ and $c^2_{(0)}$ given by 
(\ref{eqn:inversebasicreltree}) depend on $\hat v_\phi$. Still, from the
expressions  for $\hat\Pi_{WW}$ and $\hat\Pi_{ZZ}$ collected in the 
Appendix A it is clear that $M^2_W$ depends on $m_t$ quadratically.
In the limit $\hat m^2_\phi\rightarrow\infty$, in which 
$\hat v_\phi$ vanishes, $s^2_{(0)}$ and $c^2_{(0)}$ approach 
their SM values and the coefficient of $m^2_t$ on the right-hand side
of this formula becomes as in the SM.

The renormalization  scale dependence of this expression can be checked
as in the case of $\rho_{\rm low}$ using the formula 
\begin{eqnarray}
{1\over s^2_{(0)}(Q)} &\approx& {1\over s^2_{(0)}(Q^\prime)}
\left\{1+{c^2_{(0)}\over c^2_{(0)}-s^2_{(0)}}
{4\sqrt2G_F\over1-4\sqrt2G_F\hat v^2_\phi}
\dot v^2_\phi\ln{Q\over Q^\prime}\right\}
\end{eqnarray}

Similarly as for $\rho_{\rm low}$, it can be checked that in the limit 
$\hat m_\phi\rightarrow\infty$ the 1PI contributions of extra scalars
decouple from $M_W$, but the tadpoles do not cancel out and generically 
there is no decoupling
if the dimensionful coupling $\hat \mu$ in (\ref{eqn:potential}) grows
along with $\hat m_\phi$. Compared to the SM, the one-loop prediction of the 
triplet model for $M_W$ is in this limit modified only by the radiatively 
generated nonzero VEV of the triplet.

Of course, if $\hat \mu$ is kept fixed all extra contributions to $M_W$
disappear in the limit $\hat m_\phi\rightarrow\infty$. This limit is 
useful to elucidate the relation of the results obtained in \cite{DAWSON}
for $M_W$ to the SM prediction. The renormalization scheme of \cite{DAWSON}
is based on four input observables: $\alpha_{\rm EM}$, $G_F$, $M_Z$ and 
$\sin^2\theta^{\rm eff}_\ell$ (the last quantity is defined by the coupling
of on-shell $Z^0$ to on-shell charged lepton-antilepton pair).
The same scheme have been also used earlier in \cite{BLHO,LYNA}.

At one loop the basic formulae in the scheme of refs. \cite{BLHO,LYNA,DAWSON}
read
\begin{eqnarray}
M^2_Z&=&{1\over4}{\hat e^2\over\hat s^2\hat c^2}\hat v_H^2+\hat\Pi_{ZZ}(M^2_Z)
\nonumber\\
\alpha_{\rm EM} &=& {\hat e^2\over4\pi}
\left(1+\tilde\Pi_\gamma(0)-{\hat\alpha\over\pi}\ln
{\hat M^2_W\over Q^2}\right)\label{eqn:4inputbasicrel}\\
\sqrt2 G_F&=&{ \hat e^2\over 4\hat s^2\hat M^2_W}
={1\over\hat v^2_H+4\hat  v^2_\phi}\left(1+\Delta_G\right)\nonumber\\
\sin^2\theta^{\rm eff}_\ell&=&\hat s^2\left(1+\Delta_{s^2}\right)\nonumber
\end{eqnarray}
with $\Delta_G$ given in (\ref{eqn:DeltaG}) and 
\begin{eqnarray}
\Delta_{s^2}=(1-2\hat s^2)\left(\hat\Sigma_{VR}-\hat\Sigma_{VL}\right)
-{\hat c\over\hat s}{\hat\Pi_{Z\gamma}(M^2_Z)\over M^2_Z}
-{\hat c\over\hat e\hat s}\hat F_R-{2\hat s\hat c\over\hat e}
\left(\hat F_L-\hat F_R\right)
\end{eqnarray}
where $\hat F_L$, $\hat F_R$ are the 1PI corrections to the $Z^0\ell\bar\ell$
vertex and $\hat\Sigma_{VL}$, $\hat\Sigma_{VR}$ are the vector parts of the
charged lepton self energies.
Solving the relations (\ref{eqn:4inputbasicrel}) to one loop accuracy gives 
\begin{eqnarray}
\hat\alpha&=&\alpha_{\rm EM}\left(1-\tilde\Pi_\gamma(0)+
{\alpha_{\rm EM}\over\pi}\ln{(M^2_W)_{(0)}\over Q^2}\right)\nonumber\\
\hat s^2&=&\sin^2\theta^{\rm eff}_\ell\left(1-\Delta_{s^2}\right)
\label{eqn:inverserelationsfor4}\\
\hat v^2_H&=&{\sin^2\theta^{\rm eff}_\ell\cos^2\theta^{\rm eff}_\ell\over
\pi\alpha_{\rm EM}}M^2_Z\left\{
1+\tilde\Pi_\gamma(0)-
{\alpha_{\rm EM}\over\pi}\ln{(M^2_W)_{(0)}\over Q^2}\right.\nonumber\\
&&\phantom{aaaaaaaaaaaaaaaaa}\left.
-\left(1-{\sin^2\theta^{\rm eff}_\ell\over\cos^2\theta^{\rm eff}_\ell}\right)
\Delta_{s^2}-{\hat\Pi_{ZZ}(M^2_Z)\over M^2_Z}\right\}\nonumber\\
4\hat v^2_\phi&=&{1\over\sqrt2G_F}(1+\Delta_G)
-{\sin^2\theta^{\rm eff}_\ell\cos^2\theta^{\rm eff}_\ell\over
\pi\alpha_{\rm EM}}M^2_Z\left\{
1+\tilde\Pi_\gamma(0)\matrix{\phantom{a}\cr\phantom{a}}
\right.\nonumber\\
&&\left.\phantom{aaaaaaaaaaaa}
-{\alpha_{\rm EM}\over\pi}\ln{(M^2_W)_{(0)}\over Q^2}
-\left(1-{\sin^2\theta^{\rm eff}_\ell\over\cos^2\theta^{\rm eff}_\ell}\right)
\Delta_{s^2}-{\hat\Pi_{ZZ}(M^2_Z)\over M^2_Z}\right\}\nonumber
\end{eqnarray}
where now 
$(M^2_W)_{(0)}=\pi\alpha_{\rm EM}/\sqrt2G_F\sin^2\theta^{\rm eff}_\ell$.
In this scheme the one-loop formula for $M^2_W$ reads
\begin{eqnarray}
M^2_W={\pi\alpha_{\rm EM}\over\sqrt2G_F\sin^2\theta^{\rm eff}_\ell}
\left\{1-\tilde\Pi_\gamma(0)+
{\alpha_{\rm EM}\over\pi}\ln{\hat M^2_W\over Q^2}
+\Delta_{s^2}+\Delta_G+{\hat\Pi_{WW}(M^2_W)\over(M^2_W)_{(0)}}\right\}
\end{eqnarray}
It has formally the same form as the formula obtained in the SM with 
$\alpha_{\rm EM}$, $G_F$ and $\sin^2\theta^{\rm eff}_\ell$ (instead of $M_Z$)
taken for the input observables. It is also clear that tadpole contribution 
cancels out between $\hat\Pi_{WW}(M^2_W)/(M^2_W)_{(0)}$ and $\Delta_G$. The 
dependence on the top quark mass is only logarithmic: quadratic dependence
cancels out between $\hat\Pi_{WW}(M^2_W)/(M^2_W)_{(0)}$ and 
$-\hat\Pi_{WW}(0)/(M^2_W)_{(0)}$ in $\Delta_G$.
The difference with the SM is of course that in 
$\Pi_{WW}$, $\Delta_{s^2}$, $\Delta_G$ and in $\tilde\Pi_\gamma(0)$ there are 
contributions of the extended scalar sector. They would disappear 
(cancel out) in the limit $s_\gamma,s_\delta\rightarrow0$, 
$\hat M_{K^0},\hat M_{H^+}\rightarrow\hat  m_\phi$ and one would then get the 
expression for $M_W$ which is identical to the SM result in the scheme based
on $\alpha_{\rm EM}$, $G_F$ and $ \sin^2\theta^{\rm eff}_\ell$,
in which $M_W^2$ also depends on $m_t$ only logarithmically. However, in
the renormalization scheme based on $M_Z$, $\alpha_{\rm EM}$, $G_F$ and 
$\sin^2\theta^{\rm eff}_\ell$ as input observables this limit cannot be 
freely taken: while in terms of renormalized $\overline{\rm MS}$ 
parameters the limit $\hat m_\phi\rightarrow\infty$, 
$\hat M_{K^0},\hat M_{H^+}\rightarrow\hat m_\phi$ formally
entails $s_\gamma,s_\delta\rightarrow2\hat v_\phi/\hat v_H$, in the
discussed  scheme
$\hat v_\phi$ and $\hat v_H$ are in fact determined by the  equations 
(\ref{eqn:inverserelationsfor4}) and their correlation with the magnitude
of $\hat m_\phi$ is lost. In other words, in this scheme, 
whether $\hat v_\phi\rightarrow0$ and whether this limit corresponds to
the limits $\hat M_{K^0},\hat M_{H^+}\rightarrow\hat  m_\phi$ and
$s_\gamma,s_\delta\rightarrow0$ is dictated by the data and the SM
contributions to observables and not by theoretical considerations
(as is possible in our scheme).

One can wonder however, how decoupling of the triplet degrees of freedom
in the  limit, $\hat m_\phi\rightarrow\infty$, $\hat \mu$ fixed
could manifest itself in the scheme based on four input observables.
In particular one can wonder how the celebrated SM quadratic 
dependence of the $M_W\leftrightarrow M_Z$ interrelation on $m_t$ 
would be recovered?
To discuss this it is easier to imagine for a while that the four
experimental input data $\alpha_{\rm EM}$, $G_F$, $M_Z$ and 
$\sin^2\theta^{\rm eff}_\ell$
can be varied freely and that they are such that (to one loop 
accuracy)
\begin{eqnarray}
{\sin^2\theta^{\rm eff}_\ell\cos^2\theta^{\rm eff}_\ell\over
\pi\alpha_{\rm EM}}M^2_Z={1\over\sqrt2G_F}
\end{eqnarray}
so that that the equation for $4\hat v_\phi$ in 
(\ref{eqn:inverserelationsfor4}) can be satisfied by $\hat v_\phi=0$ in 
the decoupling limit $\hat m_\phi\rightarrow\infty$, $\hat\mu$ fixed (all 
additional contributions of the heavy particles cancel out or vanish in the 
right hand side of the equation for $4\hat v_\phi$). Then this equation 
relates the input observable $M^2_Z$ to the three other input observables 
$\alpha_{\rm EM}$, $G_F$ and $\sin^2\theta^{\rm eff}_\ell$. It is then 
obvious that if the model is to fit the data in the decoupling limit, the
measured $M_Z^2$ wich 
allows for $\hat v_\phi\approx0$ must change quadratically with the change 
of the input value of $m_t$. For $\hat v_\phi=0$ the last equation in 
(\ref{eqn:inverserelationsfor4}) is (to one loop accuracy) equaivalent to
\begin{eqnarray}
M^2_Z={\pi\alpha_{\rm EM}\over
\sqrt2G_F\sin^2\theta^{\rm eff}_\ell\cos^2\theta^{\rm eff}_\ell}
\left\{1-\tilde\Pi_\gamma(0)+
{\alpha_{\rm EM}\over\pi}\ln{(M^2_W)_{(0)}\over Q^2}\right.
\phantom{aaaaaaaaaaa}\nonumber\\
\left.
+\left(1-{\sin^2\theta^{\rm eff}_\ell\over\cos^2\theta^{\rm eff}_\ell}\right)
\Delta_{s^2}+\Delta_G+{\hat\Pi_{ZZ}(M^2_Z)\over M^2_Z}\right\}
\end{eqnarray}
which is precisely the formula  for $M^2_Z$ in the SM renormalized 
with $\alpha_{\rm EM}$, $G_F$ and $\sin^2\theta^{\rm eff}_\ell$
as the input observables which depends on $m_t$ quadratically.

Let us also note that
if $\alpha_{\rm EM}$, $G_F$ and $\sin^2\theta$ were taken for the input
observables in the triplet model instead of $\alpha_{\rm EM}$, $G_F$ and 
$M_Z$, the tadpole contributions to the resulting one loop 
expression for $M_W$ in the triplet model would cancel out and the formula
(\ref{eqn:Wmass}) would reduce to the SM expression in the limit 
$\hat m_\phi\rightarrow\infty$, even 
for $\hat \mu\propto\hat m_\phi$.  This can be easily 
understood: the only dangerous effect (which spoils decoupling e.g. 
in $\rho_{\rm low}$) are those which can be interpreted as corrections
to the tree level VEV $\hat v_\phi$. Since the tree level expression for 
$M_W$ in the triplet model with $\alpha_{\rm EM}$, $G_F$ and $\sin^2\theta$
as the input observables  is independent of $\hat v_\phi$, the one loop 
tadpole contributions  must cancel out. In this scheme tadpoles, which 
for $\mu\propto m_\phi$ would not decouple, would
enter $M_Z$ for which the tree level expression would 
depend on $\hat v_\phi$.

Finally is also worthwhile discussing calculation of $\rho_{\rm low}$, $M_W^2$ 
and other electroweak observables in a renormalization scheme using as the 
input observables $\alpha_{\rm EM}$, $G_F$, $M_Z$ and in addition one 
``high energy'' observable, which is absent in the SM. For example for the 
additional observable one could take the on shell value of the formfactor 
$\Lambda$ proportional to $g_{\mu\nu}$ in the $H^+$ coupling to the $Z^0W^+$ 
pair. Following the general procedure one would then express $\hat\alpha$,
$\hat s^2$, $\hat v_H$ and $\hat v_\phi$ in terms of 
$\alpha_{\rm EM}$, $G_F$, $M_Z$ and $\Lambda$. At the tree level one would
then have
\begin{eqnarray}
\hat\alpha&=&\alpha_{\rm EM}\nonumber\\
\hat s^2&=&{1\over2}\left\{1+{\Lambda^2\over4\sqrt2G_FM^4_Z}-
\sqrt{\left(1+{\Lambda^2\over4\sqrt2G_FM^4_Z}\right)^2
-{4\pi\alpha_{\rm EM}\over\sqrt2G_FM^2_Z}}\right\}\nonumber\\
\hat v_H^2&=&{1\over\sqrt2G_F}\left(1-{\Lambda^2\over4\pi\alpha_{\rm EM}M^2_Z}
\hat s^2\right)\nonumber\\
\hat v_\phi^2&=&{1\over4\sqrt2G_F}{\Lambda^2\over4\pi\alpha_{\rm EM}M^2_Z}
\hat s^2\label{eqn:anotherscheme} 
\end{eqnarray}
The decoupling limit would then correspond to taking $\Lambda\rightarrow0$
(this is possible, as $\Lambda$ is not fixed by experiment yet).

Formulae necessary to express $\hat\alpha$, $\hat s^2$, $\hat v_H^2$ and
$\hat v_\phi^2$ to one loop can be obtained by substituting in
(\ref{eqn:anotherscheme})
\begin{eqnarray}
\alpha_{\rm EM}&\rightarrow&\alpha_{\rm EM}
\left(1-\tilde\Pi_\gamma(0)+{\alpha_{\rm EM}\over\pi}
\ln{(M^2_W)_{(0)}\over Q^2}\right)\nonumber\\
{1\over\sqrt2G_F}&\rightarrow&{1\over\sqrt2G_F}\left(1+\Delta_G\right)
\nonumber\\
M_Z^2&\rightarrow& M_Z^2\left(1-{\hat\Pi_{ZZ}(M^2_Z)\over M^2_Z}\right)
\nonumber\\
\Lambda&\rightarrow&\Lambda-\delta\Lambda\nonumber
\end{eqnarray}
(where $\delta\Lambda$ is a one loop correction to the on shell $H^+Z^0W^+$
vertex) and expanding the resulting expressions appropriately.  

Since in such a scheme the tree level expression for electroweak observables 
do not depend on $\hat v_\phi$, the tadpole contributions must cancel
out (just as they do in the SM). Nondecoupling of the 
Higgs triplet effects would then manifest itself through the corrections
$\delta\Lambda$. Indeed, the $g_{\mu\nu}$ formfactor of the $H^+Z^0W^+$
coupling receives, among others, a contribution from the one loop diagram
with $H^+G^+_WH^0$ coupling and $G^+_W$, $H^0$ and $Z^0$ circulating in
the loop. Since the particles in the loop are light, the loop integral 
is not suppressed by any heavy mass factor. Moreover, 
as is easy to check, the $H^+G^+_WH^0$ coupling is proportional
to $\mu$ and in fact grows in the limit $\mu\sim m_\phi\rightarrow\infty$.
In electroweak observables the correction $\delta\Lambda$ is always
multiplied by $\Lambda$. Now, for $\mu\propto m_\phi$ and
fixed values of the dimensionless Higgs potential couplings, the limit 
$M_{K^0}\sim M_{H^+}\rightarrow m_\phi\rightarrow\infty$ requires that
$\Lambda$ vanishes only as $1/\mu$. Hence, $\propto\Lambda\delta\Lambda$
contribution to electroweak observables does not disappear and the
decoupling is violated. One should stress however, that introducing 
an additional ``high energy'' observable like $\Lambda$ makes the
analysis of the decoupling much more complicated not only from the point
of view of practial calculations but also conceptually.

\section{$S$, $T$, $U$ parameters and other issues}  

Our calculation carries also an important message for the calculations 
of the $S$, $T$ and $U$ parameters introduced in \cite{PETA} and widely 
used to constrain extensions of the SM. (Application of these parameters to 
the model considered in this paper can be found in \cite{FOROWH}.
Applications of the parameters $S$, $T$, $U$ to models with triplets have 
been also considered in \cite{STR}) Using these 
parameters implicitly assumes working in the renormalization scheme 
defined in section \ref{sec:scheme}, with $\alpha_{\rm EM}$, $G_F$
and $M_Z$ used as the only input observables and treating other parameters
of the tested model as renormalized running parameters. 

At one loop the
expressions for these parameters in the triplet model read \cite{FOROWH}
\begin{eqnarray}
\alpha_{\rm EM}T&=&{\hat\Pi_{WW}^{\rm new}(0)\over (c^2
M^2_W)_{(0)\rm SM}}-{\hat\Pi_{ZZ}^{\rm new}(0)\over M^2_Z}
+4\sqrt2 G_F\hat v^2_\phi\nonumber\\
\alpha_{\rm EM}S&=&4(s^2c^2)_{(0)\rm SM}\left\{
\hat\Pi_{ZZ}^{\prime~\rm new}-\left({c^2-s^2\over cs}\right)_{(0)\rm SM}
\hat\Pi_{Z\gamma}^{\prime~\rm new}-
\tilde\Pi_\gamma(0)^{\rm new}\right\}\label{eqn:STUdefs}\\
\alpha_{\rm EM}U&=&4s^2_{(0)\rm SM}\left\{\hat\Pi_{WW}^{\prime~\rm new}
-c^2_{(0)\rm SM}\hat\Pi_{ZZ}^{\prime~\rm new}
-2(sc)_{(0)\rm SM}\hat\Pi_{Z\gamma}^{\prime~\rm new}-s^2_{(0)\rm SM}
\tilde\Pi_\gamma(0)^{\rm new}\right\}\nonumber
\end{eqnarray}
where $\hat\Pi(q^2)=\hat\Pi(0)+q^2\hat\Pi^\prime$. Those corrections to
electroweak observables that can be interpreted as corrections to the 
gauge boson propagators can be expressed in terms of these parameters.
For example, in terms of $S$, $T$ and $U$ the corrections to the $W$ boson 
mass due to the triplet extension of the SM read
\begin{eqnarray}
\delta M_W^2=(M^2_W)_{(0)\rm SM}\left\{
\left({c^2\over c^2-s^2}\right)_{(0)\rm SM}\alpha_{\rm EM}T
+{1\over4s^2_{(0)\rm SM}}\alpha_{\rm EM}U
-{1\over2}\left({1\over c^2-s^2}\right)_{(0)\rm SM}\alpha_{\rm EM}S\right\}
\nonumber
\end{eqnarray}
This agrees with the full one loop corrections to $M_W$ (\ref{eqn:Wmass})
if one sets $v^2_\phi=0$ in the one loop part of (\ref{eqn:Wmass}),
neglects the (``nonoblique'') box and vertex corrections and expands
$s_{(0)}^2$ in the prefactor of (\ref{eqn:Wmass}) to first order in
$\hat v^2_\phi$ (this contribution is accounted by the term
$4\sqrt2G_F\hat v^2_\phi$ in $T$ in (\ref{eqn:STUdefs})).
It is therefore clear 
that the expression for $T$ should also include the tadpole contribution. 
Neglecting tadpoles in $T$ is equivalent to the (tacit) assumption that the 
the model parameters are taken at the renormalization scale $Q$ for which
tadpole contribution to $T$ happens to vanish and that it is just at this 
scale $Q$ that $v_\phi$ in the tree level term in $T$ is small. 

In general however, in all SM extensions, in which the tree level expressions
for electroweak observables depend on VEVs of additional Higgs bosons
tadpoles must be included in $T$. The minimal supersymmetric extension
(the MSSM) is special here because 
because the tree level masses of the gauge bosons depend on the
same combination of the coresponding two VEVs.

Finally let us notice that Grand Unified Theories (GUTs) generically give 
rise to $SU(2)$ triplets which are assumed to have mass parameters 
$m_\phi\sim M_{\rm GUT}$. Since in GUTs the analog of the parameter $\mu$
is also typically of the same order (it arises from a GUT gauge 
symmetry breaking VEV) nonsupersymmetric GUTs generically suffer from the 
problem of nondecoupling of $SU(2)$ triplets (the problem of justifying 
vanishing of their effects adds to the standard hierarchy problem of 
such models). \\
{\bf to ni\.zej pewnie trzeba skasowa\'c}
The problem does not arise in supersymmetric GUTs 
because there fermion contributions to tadpoles cancel against bosonic
ones in the limit of exact supersymmetry. Therefore, in realistic 
models tadpoles are suppressed by $m_{\rm soft}/M_{\rm GUT}$ where 
$m_{\rm soft}\sim{\cal O}(1$~TeV) is a typical soft supersymmetry breaking 
scale.

\section{Discussion}

We have applied the renormalization scheme based on three input observables
$\alpha_{\rm EM}$, $G_F$ and $M_Z$ to the extension of the standard model 
with a Higgs field transforming as an $SU(2)$ triplet. As we have
argued, such a scheme allows for straightforward investigation of the
question of Appelquist-Carrazzone decoupling of additional heavy particles. 
We have explicitly shown that in the model with an $Y=0$ triplet the 
decoupling does not hold if the dimensionful trilinear coupling grows
along with the triplet mass parameter. Our approach allowed us to attribute 
this effect to a nonzero triplet VEV generated by radiative corrections
for nonzero VEV of the Higgs doublet. 
We have also checked that similar
nondecoupling of heavy particle effects is present in models with $Y=\pm1$
triplets which arise in Littlest Higgs models \cite{LH} or in models 
aiming at protecting primordial baryon asymmetry \cite{HA}. At one loop
the effects of the heavy triplet in electroweak observables 
can be negligible only for severely tuned parameters of the model.
It appears, however, unlikely that they can be eliminated in this way
from all electroweak observables in higher orders.

\vskip1.0cm
\section*{Acknowledgments}

We would like to thank G. Senjanovic and J. Mourad for constructive
discussion which led us to inclusion of the \ref{app:nondecoupling}.
P.H.Ch. would like to thank the CERN Theory Group for hospitality.
P.H.Ch., and S.P. were partially supported by the European Community 
Contract MRTN-CT-2004-503369 for years 2004--2008 and by the Polish KBN 
grant 1 P03B 099 29 for years 2005--2007. 

\newpage
\renewcommand{\thesection}{Appendix~\Alph{section}}
\renewcommand{\theequation}{\Alph{section}.\arabic{equation}}
\setcounter{section}{0}

\section{Useful formulae}
\label{app:formulae}
\setcounter{equation}{0}

\noindent {\it Relations for scalar masses.}
From (\ref{eqn:massmatrices})
and (\ref{eqn:neutralHiggsrot}) the following useful relations can be derived:
\begin{eqnarray}
c^2_\gamma M^2_{H^0}+s^2_\gamma M^2_{K^0}&=&{1\over2}\lambda_Hv^2_H
\nonumber\\
s^2_\gamma M^2_{H^0}+c^2_\gamma M^2_{K^0}&=&
{1\over2}\lambda_\phi v^2_\phi+{v^2_H\over4}{\mu\over v_\phi}
\label{eqn:directrelations}\\
c_\gamma s_\gamma \left(M^2_{H^0}-M^2_{K^0}\right)&=&
v_H(\kappa v_\phi-{1\over2}\mu)\nonumber
\end{eqnarray}
and
\begin{eqnarray}
{s^2_\gamma\over M^2_{H^0}}+{c^2_\gamma\over M^2_{K^0}}&=&
{{1\over2}\lambda_Hv^2_H\over{\rm DET}}\nonumber\\
{c^2_\gamma\over M^2_{H^0}}+{s^2_\gamma\over M^2_{K^0}}&=&
{{1\over2}\lambda_\phi v^2_\phi+{v^2_H\over4}{\mu\over v_\phi}
\over{\rm DET}}
\label{eqn:inverseinverserelations}\\
s_\gamma c_\gamma
\left({1\over M^2_{H^0}}-{1\over M^2_{K^0}}\right)&=&
{-v_H(\kappa v_\phi-{1\over2}\mu)\over{\rm DET}}\nonumber
\end{eqnarray}
Where
\begin{eqnarray}
{\rm DET}\equiv M^2_{H^0}M^2_{K^0}
={1\over4}v^2_Hv^2_\phi\left[
\lambda_H\lambda_\phi 
+{\lambda_H\over2}{v^2_H\mu\over v_\phi^3}-4\kappa^2
+4{\kappa\mu\over v_\phi}-{\mu^2\over v_\phi^2}\right]\label{eqn:DET}
\end{eqnarray}
\vskip0.2cm

\noindent {\it Box and vertex correction contributions to $\Delta_G$
(\ref{eqn:DeltaG})}. They can be calculated as in the SM (except 
that the relation $\hat M^2_W=\hat c^2\hat M^2_Z$ cannot be used).
One then finds
\begin{eqnarray}
B_{\rm boxes}+2\hat\Lambda+\hat\Sigma_{e L}
+\hat\Sigma_{\nu L}=
{\hat\alpha\over4\pi\hat s^2}\left\{
-4\eta_{\rm div}-4\ln{\hat M^2_Z\over Q^2}+6
\phantom{aaaaaaaaaaaaa}\right.\nonumber\\
\left.
+\left({1\over2}-{5\over2}\hat s^2+
{7-14\hat s^2+10\hat s^4\over4\hat s^2}\left[{\hat s^2\over\hat c^2}
{\hat M^2_W\over\hat M^2_Z-\hat M^2_W}\right]
\right)\ln{\hat M^2_W\over\hat M^2_Z}\right\}\label{eqn:boxesandvertices}
\end{eqnarray}
For $\hat M^2_W=\hat c^2\hat M^2_Z$ this reduces to the SM result.
\vskip0.2cm

\noindent {\it Gauge boson self energies}.
For  $\Pi_{Z\gamma}(q^2)$ in units $\hat e^2/16\pi^2\hat s\hat c$ (in the
Feynman gauge) we get
\begin{eqnarray}
&&{1\over2}\sum_fN_c^{(f)}|Q_f|(1-4|Q_f|\hat s^2)
\left[4\tilde A(q^2,\hat m_f,\hat m_f)
+q^2~b_0(q^2,\hat m_f,\hat m_f)\right]\nonumber\\
&&-\hat c^2\left[8\tilde A(q^2,\hat M_W,\hat M_W)
+(4q^2+2\hat M^2_W)b_0(q^2,\hat M_W,\hat M_W)-{2\over3}q^2\right]
\nonumber\\
&&-{1\over2}\hat e^2v_H^2\left(1-
{\hat c^2\over\hat s^2}{4v^2_\phi\over v^2_H}\right)
b_0(q^2,\hat M_W,\hat M_W)\phantom{aaaa}\nonumber\\
&&-2\left((\hat c^2-\hat s^2)c^2_\delta
+2\hat c^2~s^2_\delta\right)\tilde A(q^2,\hat M_W,\hat M_W)
\nonumber\\
&&-2\left((\hat c^2-\hat s^2)s^2_\delta
+2\hat c^2~c^2_\delta\right)\tilde A(q^2,\hat M_{H^+},\hat M_{H^+})
\nonumber
\end{eqnarray}

For the 1PI contribution to $\hat\Pi_{ZZ}(q^2)$ the one loop formula 
(in units $\hat e^2/16\pi^2\hat s^2\hat c^2$) is
\begin{eqnarray}
&&{1\over2}\sum_fN_{c_f}a_+^f\left[2\tilde A(q^2,\hat m_f,\hat m_f)
+({q^2\over2}-\hat m^2_f)b_0(q^2,\hat m_f,\hat m_f)\right]\nonumber\\
&+&{1\over2}\sum_fN_{c_f}a_-^fm^2_fb_0(q^2,\hat m_f,\hat m_f)
+3\left[\tilde A(q^2,0,0)+{q^2\over4}b_0(q^2,0,0)\right]\nonumber\\
&-&\hat c^4\left[8\tilde A(q^2,\hat M_W,\hat M_W)
+(4q^2+2\hat M^2_W)b_0(q^2,\hat M_W,\hat M_W)
-{2\over3}q^2\right]\nonumber\\
&-&\left[(\hat c^2-\hat s^2)c_\delta^2+2\hat c^2s_\delta^2\right]^2
\tilde A(q^2,\hat M_W,\hat M_W)\nonumber\\
&-&\left[(\hat c^2-\hat s^2)s_\delta^2+2\hat c^2c_\delta^2\right]^2
\tilde A(q^2,\hat M_{H^+},\hat M_{H^+})\nonumber\\
&-&2c_\delta^2s_\delta^2
\tilde A(q^2,\hat M_W,\hat M_{H^+})\nonumber\\
&-&c_\gamma^2\tilde A(q^2,\hat M_Z,\hat M_{H^0}) \label{eqn:PIZZexpression}\\
&-&s_\gamma^2\tilde A(q^2,\hat M_Z,\hat M_{K^0}) \nonumber\\
&+&2\hat e^2\hat c^2\left({1\over2}v_H{\hat s\over\hat c}c_\delta
-v_\phi{\hat c\over\hat s}s_\delta\right)^2b_0(q^2,\hat M_W,\hat M_W)
\nonumber\\
&+&2\hat e^2\hat c^2\left({1\over2}v_H{\hat s\over\hat c}s_\delta
+v_\phi{\hat c\over\hat s}c_\delta\right)^2b_0(q^2,\hat M_W,\hat M_{H^+})
\nonumber\\
&+&\hat M^2_Z c^2_\gamma b_0(q^2,\hat M_Z,\hat M_{H^0})\nonumber\\
&+&\hat M^2_Z s^2_\gamma b_0(q^2,\hat M_Z,\hat M_{K^0})\nonumber
\end{eqnarray}
where the sums in the two first terms extend to all fermions except neutrinos
(which are accounted for by the third term),
$a_+^f=1-4|Q_f|\hat s^2+8|Q_f|^2\hat s^4$, 
$a_-^f= -4|Q_f|\hat s^2+8|Q_f|^2\hat s^4$.

Finally, in units $\hat e^2/16\pi^2\hat s^2$, for the 1PI part of 
$\hat\Pi_{WW}(q^2)$ we have
\begin{eqnarray}
&&{1\over2}\sum_{k=1}^3
\left[4\tilde A(q^2,\hat m_{e_k},0)+(q^2-\hat m^2_{e_k})
b_0(q^2,\hat m_{e_k},0)\right]\nonumber\\
&+&{3\over2}\sum_{k,l=1}^3|V_{CKM}^{kl}|^2
\left[4\tilde A(q^2,\hat m_{u_k},\hat m_{d_l})
+(q^2-\hat m^2_{u_k}-\hat m^2_{d_l})b_0(q^2,\hat m_{u_k},\hat m_{d_l})\right]
\nonumber\\
&-&\hat s^2
\left[8\tilde A(q^2,\hat M_W,0)+(4q^2+\hat M^2_W)b_0(q^2,\hat M_W,0)
-{2\over3}q^2\right]
\nonumber\\
&-&\hat c^2\left[8\tilde A(q^2,\hat M_W,\hat M_Z)
+(4q^2+\hat M^2_W+\hat M^2_Z)b_0(q^2,\hat M_W,\hat M_Z)
-{2\over3}q^2\right]\nonumber\\
&-&4\left(s_\gamma s_\delta+{1\over2}c_\gamma c_\delta\right)^2
\tilde A(q^2,\hat M_W,\hat M_{H^0})\nonumber\\
&-&4\left(-s_\gamma c_\delta+{1\over2}c_\gamma s_\delta\right)^2
\tilde A(q^2,\hat M_{H^+},\hat M_{H^0})\nonumber\\
&-&4\left(c_\gamma s_\delta-{1\over2}s_\gamma c_\delta\right)^2
\tilde A(q^2,\hat M_W,\hat M_{K^0})\nonumber\\
&-&4\left(-c_\gamma c_\delta-{1\over2}s_\gamma s_\delta\right)^2
\tilde A(q^2,\hat M_{H^+},\hat M_{K^0})\nonumber\\
&-&c_\delta^2\tilde A(q^2,\hat M_W,\hat M_Z)\label{eqn:PIWWexpression}\\
&-&s_\delta^2\tilde A(q^2,\hat M_{H^+},\hat M_Z)\nonumber\\
&+&\hat e^2\left({1\over2}v_Hc_\delta+v_\phi s_\delta\right)^2
b_0(q^2,\hat M_W,0)\nonumber\\
&+&\hat e^2\left({1\over2}v_H{\hat s\over\hat c}c_\delta
-v_\phi{\hat c\over\hat s} s_\delta\right)^2
b_0(q^2,\hat M_W,\hat M_Z)\nonumber\\
&+&\hat e^2\left({1\over2}v_H{\hat s\over\hat c}s_\delta
+v_\phi{\hat c\over\hat s} c_\delta\right)^2
b_0(q^2,\hat M_{H^+},\hat M_Z)\nonumber\\
&+&{\hat e^2\over\hat s^2}\left({1\over2}v_Hc_\gamma+2v_\phi s_\gamma\right)^2
b_0(q^2,\hat M_W,\hat M_{H^0})\nonumber\\
&+&{\hat e^2\over\hat s^2}
\left(-{1\over2}v_Hs_\gamma+2v_\phi c_\gamma\right)^2
b_0(q^2,\hat M_W,\hat M_{K^0})\nonumber
\end{eqnarray}
\vskip0.2cm

\noindent {\it Trilinear couplings of $H^0$ and $K^0$ relevant for tadpoles}. 
The couplings are written as e.g. 
${\cal L}\supset-\lambda^0_{HGG}H^0G^0G^0-\lambda^\pm_{HGG}H^0G^+_WG^-_W$ etc. 
and the factors $\lambda$ read
\begin{eqnarray}
\lambda^0_{HGG}&=&{1\over2}\left[{1\over2}\lambda_Hv_Hc_\gamma
+\left(\kappa v_\phi-{1\over2}\mu\right)s_\gamma\right]
\nonumber\\
\lambda^0_{KGG}&=&{1\over2}\left[-{1\over2}\lambda_Hv_Hs_\gamma
+\left(\kappa v_\phi-{1\over2}\mu\right)c_\gamma\right]
\label{eqn:scalartrilinear}\\
\lambda^0_{HHH}&=&{1\over2}\left[
 {1\over2}\lambda_H    v_H              c_\gamma^3           +
 {1\over2}\lambda_\phi v_\phi                      s_\gamma^3+
          \kappa       v_H              c_\gamma   s_\gamma^2+
\left(\kappa v_\phi-{1\over2}\mu\right) c_\gamma^2 s_\gamma\right]
\nonumber\\
\lambda^0_{KHH}&=&{1\over2}\left[       
-{3\over2}\lambda_H    v_H              c_\gamma^2 s_\gamma  +
 {3\over2}\lambda_\phi v_\phi           c_\gamma   s_\gamma^2-
          \kappa       v_H             (s_\gamma^3-2c_\gamma^2 s_\gamma)+
\left(\kappa v_\phi-{1\over2}\mu\right)(c_\gamma^3-2c_\gamma s_\gamma^2)\right]
\nonumber\\
\lambda^0_{HKK}&=&{1\over2}\left[
 {3\over2}\lambda_H    v_H              c_\gamma   s_\gamma^2+
 {3\over2}\lambda_\phi v_\phi           c_\gamma^2 s_\gamma  +
          \kappa       v_H             (c_\gamma^3-2c_\gamma s_\gamma^2)+
\left(\kappa v_\phi-{1\over2}\mu\right)(s_\gamma^3-2s_\gamma c_\gamma^2)\right]
\nonumber\\
\lambda^0_{KKK}&=&{1\over2}\left[
-{1\over2}\lambda_H    v_H                         s_\gamma^3+
 {1\over2}\lambda_\phi v_\phi           c_\gamma^3           -
          \kappa       v_H              c_\gamma^2 s_\gamma  +
\left(\kappa v_\phi-{1\over2}\mu\right) c_\gamma   s_\gamma^2\right]
\nonumber\\
\lambda^\pm_{HGG}&=&
{1\over2}\lambda_H v_H c_\gamma c^2_\delta
+{1\over2}\lambda_\phi v_\phi s_\gamma s^2_\delta
+\kappa v_Hc_\gamma s^2_\delta
+\left(\kappa v_\phi+{1\over2}\mu\right) s_\gamma c^2_\delta
-\mu  c_\gamma s_\delta c_\delta
\nonumber\\
\lambda^\pm_{KGG}&=&
-{1\over2}\lambda_H v_H s_\gamma c^2_\delta
+{1\over2}\lambda_\phi v_\phi c_\gamma s^2_\delta
-\kappa v_H s_\gamma s^2_\delta
+\left(\kappa v_\phi+{1\over2}\mu\right) c_\gamma c^2_\delta
+\mu  s_\gamma s_\delta c_\delta
\nonumber\\
\lambda^\pm_{HHH}&=&{1\over2}\lambda_H v_H c_\gamma s^2_\delta
+{1\over2}\lambda_\phi v_\phi s_\gamma c^2_\delta
+\kappa v_Hc_\gamma c^2_\delta
+\left(\kappa v_\phi+{1\over2}\mu\right) s_\gamma s^2_\delta
+\mu  c_\gamma s_\delta c_\delta
\nonumber\\
\lambda^\pm_{KHH}&=&
-{1\over2}\lambda_H v_H s_\gamma s^2_\delta
+{1\over2}\lambda_\phi v_\phi c_\gamma c^2_\delta
-\kappa v_H s_\gamma c^2_\delta
+\left(\kappa v_\phi+{1\over2}\mu\right) c_\gamma s^2_\delta
-\mu  s_\gamma s_\delta c_\delta\nonumber\\
\lambda^\pm_{HGH}&=&{1\over2}\lambda_H v_H c_\gamma c_\delta s_\delta
-{1\over2}\lambda_\phi v_\phi s_\gamma c_\delta s_\delta
+\kappa(v_\phi s_\gamma -v_Hc_\gamma)c_\delta s_\delta
+{1\over2}\mu(c^2_\delta-s^2_\delta)c_\gamma\nonumber\\
\lambda^\pm_{KGH}&=&-{1\over2}\lambda_H v_H s_\gamma c_\delta s_\delta
-{1\over2}\lambda_\phi v_\phi c_\gamma c_\delta s_\delta
+\kappa(v_\phi c_\gamma +v_Hs_\gamma)c_\delta s_\delta
-{1\over2}\mu(c^2_\delta-s^2_\delta)s_\gamma\nonumber
\end{eqnarray}
\newpage

\noindent {\it Renormalization group equations.} For the $Y=0$ triplet model
the renormalization group equations for the potential 
parameters are given in \cite{FOSAVEWH} and read
\begin{eqnarray}
Q{dm^2_H\over dQ}&=&\left[-\left({9\over2}g^2_2+{3\over2}g^2_y-
2\sum_fY_f^2-3\lambda_H\right)m^2_H
+3\kappa m^2_\phi+{3\over2}\mu^2\right]\nonumber\\
Q{dm_\phi^2\over dQ}&=&\left[-\left(12g^2_2-{5\over2}\lambda_\phi\right) 
m^2_\phi+4\kappa m^2_H+\mu^2\right]\label{eqn:dpodtmasy}
\end{eqnarray}
\begin{eqnarray}
Q{d\lambda_H\over dQ}&=&\left[{9\over2}g^4_2+
3g^2_2g^2_y+{3\over2}g^4_y-
\left(9g^2_2+3g^2_y-4\sum_fY^2_f\right)\lambda_H
+6\lambda_H^2 +6\kappa^2-8\sum_fY^4_f\right]
\nonumber\\
Q{d\lambda_\phi\over dQ}&=&\left[48g^4_2-24g^2_2\lambda_\phi
+{11\over2}\lambda_\phi^2+8\kappa^2\right]
\nonumber\\
Q{d\kappa\over dQ}&=&\left[6g^4_2-\left({33\over2}g^2_2+{3\over2}g^2_y
-2\sum_fY^2_f\right)\kappa
+3\lambda_H\kappa+{5\over2}\lambda_\phi\kappa+4\kappa^2\right]
\nonumber\\
Q{d\mu\over dQ}&=&\mu\left(-{21\over2}g^2_2-{3\over2}g^2_y+2\sum_fY^2_f
+\lambda_H+4\kappa\right)\label{eqn:dpodtsprzezenia}
\end{eqnarray}
The renormalization group equation for $v_\phi$ can be derived combining
the equations (\ref{eqn:minimizconds}) with (\ref{eqn:dpodtmasy}) and 
(\ref{eqn:dpodtsprzezenia}). After some algebra one gets
\begin{eqnarray}
{1\over\lambda_H}\left[\dots
\right]~\dot v^2_\phi&=&-16\sum_fN_c^{(f)}Y^4_f(\kappa v_\phi-{1\over2}\mu)
{v_H^2\over\lambda_Hv_\phi}\nonumber\\
&+&3(g_y^2+g^2_2)^2(\kappa v_\phi-{1\over2}\mu)
{v_H^2\over\lambda_Hv_\phi}\nonumber\\
&+&6g_2^4(\kappa v_\phi-{1\over2}\mu)
{v_H^2+4v_\phi^2\over\lambda_Hv_\phi}\nonumber\\
&-&12g_2^4(v_H^2+4v_\phi^2)\nonumber\\
&+&12g^2_2v_\phi^2{1\over\lambda_H}\left[\dots\right]\nonumber\\
&-&{5\over2}\lambda_\phi\mu {v_H^2\over v_\phi}
- 3\lambda_\phi^2v_\phi^2 
+ 4\lambda_H\kappa v_H^2\nonumber\\
&-& 8\mu^2
- 8\kappa^2v_H^2
- 2\lambda_H\mu {v_H^2\over v_\phi}\nonumber\\
&+& 4\kappa \mu {v_H^2\over v_\phi}
+ 6\kappa^2 \mu {v_H^2\over\lambda_H v_\phi}
- 24{\kappa^2\mu v_\phi\over\lambda_H}\nonumber\\
&+& 4\lambda_\phi {\kappa^2\over\lambda_H}v_\phi^2
+ 16\kappa^3{v_\phi^2\over\lambda_H}
+ 20\kappa {\mu^2\over\lambda_H}\nonumber\\
&-& 2\lambda_\phi\kappa\mu {v_\phi\over\lambda_H}
- 3\kappa \mu^2 {v_H^2\over\lambda_H v_\phi^2}
- 6{\mu^3\over\lambda_Hv_\phi}
\label{eqn:RGEforvphi}
\end{eqnarray}
where we have denoted by $[\dots]$ the expression 
\begin{eqnarray}
[\dots]\equiv 
\lambda_H\lambda_\phi+4{\kappa\mu\over v_\phi}
-4\kappa^2-{\mu^2\over v_\phi^2}+{\lambda_H\over2}{\mu v^2_H\over v^3_\phi}
\end{eqnarray}
proportional to DET in (\ref{eqn:DET}): DET$={1\over4}v_H^2v_\phi^2[\dots]$.
\vskip0.2cm

\noindent {\it Some loop functions.} For completeness we recall the 
definitions:
\begin{eqnarray}
16\pi^2 a(m)= m^2\left(\eta_{\rm div}-1+\ln{m^2\over\mu^2}\right)
\end{eqnarray}
\begin{eqnarray}
16\pi^2 b_0(q^2,m_1,m_2)=\eta_{\rm div}+\int_0^1dx
\ln{q^2x(x-1)+xm^2_1+(1-x)m^2_2\over\mu^2}
\end{eqnarray}
\begin{eqnarray}
\tilde A(q^2,m_1,m_2)&=& -{1\over6}a(m_1)-{1\over6}a(m_2)
+{1\over6}(m_1^2+m_2^2-{q^2\over2})~b_0(q^2,m_1,m_2) \nonumber\\
&+&{m_1^2-m_2^2\over12q^2}\left[a(m_1)-a(m_2)-(m_1^2-m_2^2)~b_0(q^2,m_1,m_2)
\right]\nonumber\\
&-&{1\over16\pi^2}{1\over6}(m_1^2+m_2^2-{q^2\over3})
\end{eqnarray}
$\tilde A(0,m_1,m_2)$ is finite and reads
\begin{eqnarray}
16\pi^2 \tilde A(0,m_1,m_2)
=-{1\over8}\left[m_1^2+m_2^2-{2m_1^2m_2^2\over m_1^2-m_2^2}
\log{m_1^2\over m_2^2}\right]\equiv-{1\over8}g(m_1,m_2)
\end{eqnarray}

\section{Nondecoupling of the Higgs triplet}
\label{app:nondecoupling}
\setcounter{equation}{0}

Since the nondecoupling of the Higgs triplet effects may have important 
consequences for model building we elucidate it here from another point
of view in a simple model with two fields only, the doublet $H$ and the
triplet $\phi$. The most general interaction potential is given by  
(\ref{eqn:potential}). It is instructive to look at this problem from the 
symmetry point of view. The original triplet model (\ref{eqn:potential}) 
has the global $SU(2)\times U(1)$ symmetry. VEVs of $H$ and $\phi$ break 
this symmetry down to $U(1)$ (identified in the electroweak model with 
the electromagnetic symmetry). After spontaneous symmetry breaking by VEVs
(\ref{eqn:VEVs}) the fields $G^0$ and $G^\pm_W$ are massless to all orders 
and their scattering amplitudes have properties specific for Goldstone bosons
(in particular they vanish at the threshold). Decoupling of the triplet would 
mean that in the limit $m^2_\phi\rightarrow\infty$ all scattering amplitudes 
of these Goldstone bosons and $H^0$ can be reproduced by starting from the 
Lagrangian containing only the doublet $H$: 
\begin{eqnarray}
{\cal L}_{\rm eff}=z_H\partial_\mu H^\dagger\partial^\mu H+
z_Hm^2_{\rm eff}H^\dagger H+z_H^2{\lambda_{\rm eff}\over4}
\left(H^\dagger H\right)^2~,
\label{eqn:Leffective}
\end{eqnarray}
where $H=(G^+_W,(v+H^0+iG^0)/\sqrt2)^T$. The factors $z_H=1+\delta z_H$, 
$m^2_{\rm eff}=m_H^2+\delta m^2_H$ and 
$\lambda_{\rm eff}=\lambda_H+\delta\lambda_H$ would be then determined 
order by order in perturbation calculus. The model (\ref{eqn:Leffective}) is 
known to posses the $SU(2)_L\times SU(2)_R$ symmetry which the $H$ VEV breaks 
down to the so-called custodial $SU(2)_V$ symmetry. The $G^0$ and $G^\pm_W$ 
amplitudes calculated in the model (\ref{eqn:Leffective}) satisfy therefore 
$SU(2)_V$ relations which in principle do not follow from the original 
theory: in the potential (\ref{eqn:potential}) the larger 
$SU(2)_L\times SU(2)_R$ symmetry is explicitly broken by the $\mu$ term. 
Hence, one can expect that when $\mu$ grows along with $m_\phi$ the effects 
of explicit $SU(2)_L\times SU(2)_R$ breaking do not disappear. If it is 
indeed the case, then no effective renormalizable Lagrangian for fields $G^0$, 
$G^\pm_W$ and $H^0$ can reproduce the amplitudes obtained from 
(\ref{eqn:potential}). This is because in the original theory 
(\ref{eqn:potential}) $G^0$, $G^\pm_W$ are true Goldstone bosons and such an 
effective Lagrangian would have to ensure their masslessness to all orders. 
This is only possible if they are Goldstone bosons also at the effective 
Lagrangian level. (\ref{eqn:Leffective}) is however the only renormalizable 
Lagrangian that ensures masslessness of $G^0$, $G^\pm_W$ but it leads to exact 
$SU(2)_V$ symmetry. This is im contrast to what one could expect from
looking at Feynman diagrams in the symmetric phase of the original theory with 
$H$ and $\phi$: one could think that even for $\mu\sim m_\phi$ the 
decoupling should hold with the tree level matching condition (corrected
successively in higher loops)
\begin{eqnarray}
\lambda_{\rm eff}=\lambda_H-{\mu^2\over2m_\phi^2}~.
\end{eqnarray}
This is indeed how the decoupling works at the tree level even in the 
broken phase because for $m_\phi\rightarrow\infty$ the tree level VEV 
$v_\phi$ vanishes irrespectively of the behaviour of $\mu$.

\begin{figure}[]
\begin{center}
\begin{picture}(250,80)(0,0)
\DashLine(20,0)(50,30){2}
\DashLine(50,30)(20,60){2}
\Line(50,30)(100,30)
\Line(80,30)(80,60)
\Vertex(80,70){12}
\Text(0,5)[]{$G^0,G^\pm_W$}
\Text(0,55)[]{$G^0,G^\pm_W$}
\Text(65,20)[]{$K^0$}
\Text(100,45)[]{$H^0,K^0$}
\Text(100,20)[]{$H^0$}
\Vertex(50,30){1}
\Vertex(80,30){1}
\DashLine(150,0)(210,30){2}
\DashLine(150,60)(180,45){2}
\Line(180,45)(210,30)
\Line(210,30)(240,30)
\Line(180,45)(205,65)
\Vertex(210,70){12}
\Vertex(180,45){1}
\Vertex(210,30){1}
\Text(140,5)[]{$G^\pm_W$}
\Text(140,55)[]{$G^\pm_W$}
\Text(190,30)[]{$H^\pm$}
\Text(240,20)[]{$H^0$}
\Text(217,48)[]{$H^0,K^0$}
\end{picture}
\end{center}
\caption{Todpole contributions to the off-shell $H^0G^0G^0$ and
$H^0G^\pm_WG^\mp_W$ vertices in the oroginal model. For $H^0G^\pm_WG^\mp_W$ 
vertex there is also the diagram with tadpoles attached to the lower
$G^\pm_W$ leg.}
\label{fig:tadpoles}
\end{figure}
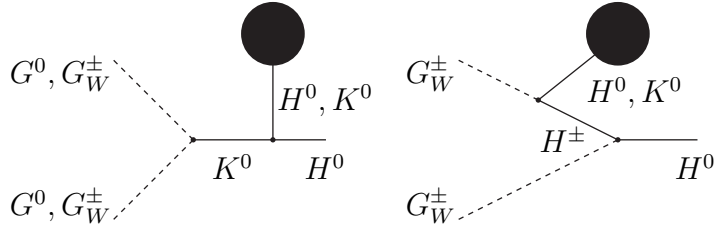

To show that for $\mu\sim m_\phi$ the Lagrangian (\ref{eqn:Leffective}) 
indeed {\it cannot} reproduce amplitudes of the original theory 
it is suffcient to point out
only one contribution which does not follow from (\ref{eqn:Leffective}).
To this end we consider one loop tadpole corrections to the $H^0G^0G^0$ and 
$H^0G^+G^-$ couplings shown in figure \ref{fig:tadpoles}. If the decoupling
holds, such contributions, although 1-particle reducible, should be 
(up to terms suppressed as $m_\phi\rightarrow\infty$) reproduced by the 
1-PI $H^0G^0G^0$ and $H^0G^+G^-$ vertices  calculated in the effective theory 
(\ref{eqn:Leffective}).
The contribution of the diagram  \ref{fig:tadpoles} reads
\begin{eqnarray}
-i2\lambda^0_{KGG}{1\over M^2_{K^0}-q^2}\left\{
2\lambda^0_{KHH}{{\cal T}_H\over M^2_{H^0}}+
2\lambda^0_{HKK}{{\cal T}_K\over M^2_{K^0}}\right\}\phantom{aaaaaaaaaaaa}
\label{eqn:nondecouplingcontrib}\\
=-i2\lambda^0_{KGG}{1\over M^2_{K^0}-q^2}\left\{\dots+
\left(\kappa v_\phi-{1\over2}\mu\right)(c^2_\gamma-s^2_\gamma)
\left({c_\gamma\over M^2_{H^0}}
{\cal T}_H-{s_\gamma\over M^2_{H^0}}{\cal T}_K\right)+\dots\right\}\nonumber
\end{eqnarray}
where we have displayed only the most important term. By using the relations
(\ref{eqn:inverseinverserelations}) it is easy to show that the leading terms
in the combination of tadpoles appearing in (\ref{eqn:nondecouplingcontrib}) 
comes from the contribution of $K^0$ and $H^+$
\begin{eqnarray}
{c_\gamma\over M^2_{H^0}}{\cal T}_H-{s_\gamma\over M^2_{H^0}}{\cal T}_K=
{1\over M^2_{H^0}M^2_{K^0}}v_H\left(\kappa m^2_\phi+{1\over2}\mu^2\right)
\left[{1\over2}a(M_{K^0})+a(M^2_{H^\pm})\right]
\end{eqnarray}
Since $a(M_{K^0})\sim a(M^2_{H^\pm})\sim m^2_\phi$ this is of order
$m^2_\phi$. Therefore, expanding (\ref{eqn:nondecouplingcontrib}) in powers
of $q^2/M^2_{K^0}$ we get for $\mu\sim m_\phi$ among the unsuppressed terms 
the contribution
\begin{eqnarray}
-i2\lambda^0_{KGG}{q^2\over M^4_{K^0}}\left(-{1\over2}\mu\right)
{1\over M^2_{H^0}M^2_{K^0}}v_H\left(\kappa m^2_\phi+{1\over2}\mu^2\right)
\left[{1\over2}a(M_{K^0})+a(M^2_{H^\pm})\right]
\end{eqnarray}
Since $2\lambda^0_{KGG}=(-{1\over2}\mu)+\dots$ we see that if $\mu\sim m_\phi$
there is a contribution to the vertex $H^0G^0G^0$ which is unsuppressed.
It is also a matter of simple analysis to see that nontadpole contributions
(as well as diagrams with tadpoles attached directly to the $H^0G^0G^0$
vertex via the $H^0H^0G^0G^0$ and $H^0K^0G^0G^0$ quartic couplings) cannot
give such a contribution. Clearly, this contribution
cannot be reproduced by the renormalizable Lagrangian (\ref{eqn:Leffective}).
Moreover, since the coupling $\lambda^\pm_{KGG}$
has the leading term proportional to $\mu$ with opposite sign compared to 
$2\lambda^0_{KGG}$, it is clear that the similar contribution to the 
$H^0G^+_WG^-_W$ vertex is different, thus breaking the custodial $SU(2)_V$ 
symmetry. Another contribution to the $H^0G^+_WG^-_W$ vertex comes from
the diagrams shown in figure \ref{fig:tadpoles}b and c.
For the leading terms they give 
\begin{eqnarray}
-i\left({1\over2}\mu\right)\left[{p_1^2\over M^4_{H^\pm}}+
{p_2^2\over M^4_{H^\pm}}\right]
\left(-{1\over2}\mu\right)
{v_H\over M^2_{H^0}M^2_{K^0}}\left(\kappa m^2_\phi+{1\over2}\mu^2\right)
\left[{1\over2}a(M_{K^0})+a(M^2_{H^\pm})\right]\nonumber\\
\end{eqnarray}
which again cannot be reproduced by the renormalizable Lagrangian.

Thus, we have shown that indeed, for $\mu\sim m_\phi$ in the original 
theory there are corrections to the $H^0G^0G^0$ and $H^0G^+_WG^-_W$ vertices 
which would require nonrenormalizable terms
\begin{eqnarray}
\Delta{\cal L}_{\rm eff}\propto {v_H\over M^2_{H^0}}
\left[{1\over2}G^0G^0(\partial^2H^0)-G^+_WG^-_W(\partial^2H^0)
+(\partial^2G^+_W)G^-_WH^0+G^+_W(\partial^2G^-_W)H^0\right]\nonumber\\
\end{eqnarray}
and which break the custodial $SU(2)_V$ symmetry.

\end{document}